\documentclass[fleqn,usenatbib]{mnras}

\usepackage{newtxtext,newtxmath}
\usepackage[T1]{fontenc}
\usepackage{ae,aecompl}
\usepackage{booktabs}

\usepackage{graphicx}	% Including figure files
\usepackage{amsmath}	% Advanced maths commands
\usepackage{amssymb}	% Extra maths symbols
\usepackage{bm}
\usepackage{threeparttable}

\newcommand{\msn}{p_{\mathrm{obs}}}
\newcommand{\inc}{\text{in}}

\title[Cosmological constraints from \textit{Planck} galaxy clusters with CMB lensing mass bias calibration]{Cosmological constraints from \textit{Planck} galaxy clusters with CMB lensing mass bias calibration}

\author[\'I. Zubeldia et al.]{
\'I\~{n}igo Zubeldia$^{1,2}$\thanks{E-mail: inigo.zubeldia@ast.cam.ac.uk}
and Anthony Challinor$^{1,2,3}$\thanks{E-mail: a.d.challinor@ast.cam.ac.uk}
\\
$^{1}$Institute of Astronomy, Madingley Road, Cambridge CB3 0HA, UK\\
$^{2}$Kavli Institute for Cosmology Cambridge, Madingley Road, Cambridge CB3 0HA, UK\\
$^{3}$DAMTP, Centre for Mathematical Sciences, Wilberforce Road, Cambridge CB3 0WA, UK
}

\date{Accepted XXX. Received YYY; in original form ZZZ}

\pubyear{2019}

\begin{document}
\label{firstpage}
\pagerange{\pageref{firstpage}--\pageref{lastpage}}
\maketitle

\begin{abstract}
We present a new cosmological analysis of the galaxy clusters in the \textit{Planck} MMF3 cosmology sample with a cosmic microwave background (CMB) lensing calibration of the cluster masses. As demonstrated by \textit{Planck},
galaxy clusters detected via the Sunyaev--Zel'dovich (SZ) effect offer a powerful way to constrain cosmological parameters such as $\Omega_{\mathrm{m}}$ and $\sigma_8$. Determining the absolute cluster mass scale is, however, difficult, and some recent calibrations have yielded cosmological constraints in apparent tension with constraints in the $\Lambda$CDM model derived from the power spectra of the primary CMB anisotropies. In order to calibrate the absolute mass scale of the full \textit{Planck} cluster sample, we remeasure the masses of all 433 clusters through their weak lensing signature in the CMB temperature anisotropies as measured by \textit{Planck}. We perform a joint Bayesian analysis of the cluster counts and masses taking as input the estimated cluster masses, SZ signal-to-noise ratios, and redshifts. Our analysis properly accounts for selection effects in the construction of the cluster sample. We find $\sigma_8 \left(\Omega_{\mathrm{m}}/0.33\right)^{0.25} = 0.765 \pm 0.035$ and $1-b_{\textnormal{SZ}} = 0.71 \pm 0.10$, where the mass bias factor $1-b_{\textnormal{SZ}}$ relates cluster mass to the SZ mass that appears in the X-ray-calibrated cluster scaling relations. We find no evidence for tension with the \textit{Planck} primary CMB constraints on $\Lambda$CDM model parameters.
\end{abstract}


%%%%%%%%%%%%%%%%%%%%%%%%%%%%%%%%%%%%%%%%%%%%%%%%%%

%%%%%%%%%%%%%%%%% BODY OF PAPER %%%%%%%%%%%%%%%%%%

%describe fiducial cosmology

\section{Introduction}
\label{sec:intro}

Galaxy clusters, the largest gravitationally-bound structures in the Universe, are powerful cosmological probes \citep{Allen2011}. In particular, their abundance as a function of mass and redshift is very sensitive to the mean matter density of the Universe, which can be parametrised by $\Omega_{\mathrm{m}}$, and to the amplitude of the matter perturbations, which can be characterised by $\sigma_8$, the root mean square of the linear density fluctuations smoothed on a scale of 8\,$h^{-1}$\,Mpc. In recent years, large Sunyaev--Zel'dovich (SZ) cluster surveys, in which clusters are detected through their thermal-SZ (tSZ) signature~\citep{Sunyaev1972}, have provided useful observations of this abundance from which cosmological constraints have been obtained~(e.g., \citealt{Staniszewski2009,Mantz2010,Hasselfield2013,Bleem2015,Planck2016xxvii,deHaan2016,Bocquet2018,Salvati2018}). SZ surveys are particularly interesting because the change in surface brightness of the cosmic microwave background (CMB) due to the SZ effect does not decrease with cluster redshift, allowing for the detection of high-redshift galaxy clusters given enough resolution. In addition, the observational selection of the cluster sample is typically straightforward to model, simplifying the extraction of cosmological information.

A crucial element in analysing the abundance of galaxy clusters (`cluster counts' analysis) is the observational determination of cluster masses. In SZ surveys, the mass of a galaxy cluster is typically estimated directly from CMB data through its mass-dependent SZ signal: the SZ signal is said to be a proxy of the galaxy cluster mass. However, the scaling relations between a cluster's mass and its SZ signal are not very well determined, and usually they need to be calibrated for each survey. This determination of cluster masses currently provides the largest source of uncertainty when obtaining cosmological information from galaxy clusters \citep{Ade2016,Pratt2019}.

The \textit{Planck} experiment detected about 1200 galaxy clusters via their SZ signature \citep{Planck2016xxvii}. A subsample of 439 such clusters, known as the MMF3 cosmology sample, was used in a counts analysis in order to constrain, amongst other parameters and models, $\Omega_{\mathrm{m}}$ and $\sigma_8$ within the context of a standard spatially-flat $\Lambda$CDM cosmology \citep{Ade2016}. This analysis proceeded as follows.
Each cluster was characterised by two observables: SZ signal-to-noise ratio, $q$, and redshift, $z$. 
The signal-to-noise ratio $q$, a proxy of the cluster mass, was measured from \textit{Planck} data for each cluster. On the other hand, the redshifts of most clusters in the sample (433) were measured by follow-up observations. These two cluster observables were then binned on a grid in the $q$--$z$ plane. The number of clusters within each cell was modelled as being independently Poisson distributed in a likelihood analysis, with the mean number of clusters being the theoretically-predicted quantities dependent on cosmology.

Theory predicts the number of clusters as a function of their redshift and true mass [we use $M_{500}$, the mass within a radius $R_{500}$, inside which the mean matter density is 500 times that of the critical density at the cluster's redshift, $\rho_c(z)$]. To connect these predictions to the expected counts in the $q$--$z$ plane, the analysis of~\citet{Ade2016} used the SZ-mass scaling relations from \citet{Arnaud2010}, with parameters calibrated with X-ray mass estimates of a subsample of clusters of the MMF3 cosmology sample (see \citealt{Planck2014} and \citealt{Planck2016xxvii} for the calibration details). However, it is known that the X-ray mass estimates are typically biased low. To account for this, a mass bias factor was introduced, such that in the X-ray-calibrated scaling relations used in the likelihood, the X-ray-derived masses were replaced by the scaled true masses,
$(1-b) M_{500}$, where $1-b$ parametrises the mass bias. There are several possible sources of this bias: it is known that X-ray cluster mass estimates are typically biased low at a significant level due to their being obtained under the assumption of 
hydrostatic equilibrium within the cluster, an assumption that can be violated in several scenarios \citep{Nagai2007,Piffaretti2008,Meneghetti2010}; there can also be observational systematic effects in the X-ray mass estimates and selection effects biasing X-ray-selected samples relative to SZ-selected samples \citep{Ade2016}.

To obtain cosmological constraints, it is necessary to calibrate the mass bias parameter
$1-b$ since this determines the overall cluster mass scale. For example, an increase in $\sigma_8$, which increases the number of clusters above a given true mass, can be offset by a reduction in $1-b$, which makes the true masses of the observed clusters larger, preserving the expected number of observed clusters. In the analysis of~\citet{Ade2016}, the mass bias was estimated from three independent calibrations. In each of these calibrations, lensing mass estimates of the clusters of a subsample of the MMF3 cosmology sample were directly compared to the corresponding masses that are obtained from the X-ray calibrated SZ-mass scaling relations if no mass bias is assumed. Lensing probes the total cluster mass distribution without relying on any of the assumptions that underlie X-ray mass estimates, and so is arguably the most robust way to measure cluster masses to date. Two of the calibrations used in~\citet{Ade2016} were from galaxy weak-lensing mass estimates of a small number of clusters in the MMF3 cosmology sample: $1 - b = 0.688 \pm 0.072$ from the Weighing the Giants (WtG,~\citealt{Linden2014}) programme; and $1-b = 0.78 \pm 0.07$ from the Canadian Cluster Comparison Project (CCCP,~\citealt{Hoekstra2015}). Both galaxy weak-lensing calibrations support a significant mass bias.
The third calibration used CMB lensing mass estimates of most of the clusters in the MMF3 cosmology sample (433) and reported $1/(1-b) = 0.99 \pm 0.19$ \citep{Ade2016}, which is consistent with their being no mass bias. These three different calibrations were then used as priors on the mass bias parameter in the cluster counts likelihood analysis. They affect the results significantly, particularly the constraints on $\sigma_8$. As shown in Fig. 7 of \citet{Ade2016}, the absence of a mass bias, a scenario suggested by the CMB lensing calibration, leads to parameter constraints in the $\Omega_{\mathrm{m}}$--$\sigma_8$ plane that are in tension with the corresponding constraints derived from the \textit{Planck} CMB power spectra in the $\Lambda$CDM model. On the other hand, the two galaxy weak-lensing mass bias calibrations alleviate the tension significantly. Model-dependent constraints on the mass bias can be obtained by combining cluster counts with the CMB power spectra. Within $\Lambda$CDM, the combination of the \textit{Planck} temperature and polarization power spectra (the \textit{Planck} 2015 TT,TE,EE+lowP likelihood) and the \textit{Planck} cluster-counts likelihood gives a posterior distribution $1-b = 0.58 \pm 0.04$~\citep{Ade2016}. This confirms that a significant mass bias is required if the $\Lambda$CDM cosmology favoured by the primary CMB anisotropies is to be consistent with the observed counts of galaxy clusters.

The mass calibration from joint analysis of the cluster counts and primary CMB power spectra was recently updated in~\citet{Planck2018} to use the \textit{Planck} 2018 TT,TE,EE+lowE+lensing likelihood (the cluster-counts likelihood was unchanged from~\citealt{Ade2016}). The main relevant change is the improved constraints on large-scale polarization due to the use of the data from the \text{Planck} High Frequency Instrument rather than the less precise data from the Low Frequency Instrument used in the 2015 likelihoods. This change favours a lower central value for the optical depth to reionization, $\tau$, with the new constraint also being around twice as precise. Lowering $\tau$ reduces the fluctuation amplitude to preserve the amplitude of the CMB spectra on smaller scales and leads to a lower predicted value of $\sigma_8$. Consistency with the observed cluster counts then requires a larger value of $1-b$ (i.e., less mass bias), with~\citet{Planck2018} reporting $1-b = 0.62 \pm 0.03$.

In this work we revisit the \textit{Planck} cluster counts analysis for the particular case of the CMB lensing mass bias calibration. After measuring again the masses of all the clusters in the \textit{Planck} CMB lensing calibration sample through their CMB lensing signature, we argue that a mass bias calibration like the one presented in \citet{Ade2016} is biased by several effects.

In order to account for these effects, we present an alternative approach in which the CMB lensing mass estimates of each cluster, along with $q$ and $z$, are directly incorporated into a likelihood that is able to constrain jointly $\Omega_{\mathrm{m}}$, $\sigma_8$, and $1 - b$ at once. We present the constraints obtained from our analysis within the $\Lambda$CDM model, finding good agreement with the constraints from the \textit{Planck} CMB power spectra.

This paper is organised as follows. First, in Section \ref{data} we detail the cluster sample used in our analysis. In Section \ref{formalism} we briefly describe the basics of CMB lensing. Next, in Section \ref{estimation} we give a detailed account of the pipeline that we follow in order to estimate the masses of the clusters in our sample via CMB lensing. In Section \ref{naive} we explain why a mass bias calibration such as the one presented in \citet{Ade2016} is, in general, biased, and in Section \ref{bayesian} we present a likelihood that, by combining the CMB lensing mass estimates of each cluster in our sample with the corresponding $q$ and $z$, has the power to constrain $\Omega_{\mathrm{m}}$, $\sigma_8$, and $1 - b$ jointly in an unbiased way. We validate this approach with simulated data in Section \ref{validation}. Our parameter results are presented and discussed in Section~\ref{results}.
% and discussed further in Section~\ref{discussion}.
We conclude in Section \ref{conclusions}. Finally, in Appendix~\ref{appendix1} a method for dealing with uncertainty in cluster centering in future analyses is proposed, and in Appendix~\ref{appendix3} the relation of our likelihood to a simple Poisson counts likelihood in the space of cluster observables is discussed.

\section{Data}\label{data}

For our cosmological analysis we use the clusters of the \textit{Planck} MMF3 cosmology sample. This sample contains a total of 439 clusters and is the sample that was used in the baseline analysis in \citet{Ade2016}. Details of the construction of this sample are given in~\citet{Planck2016xxvii} and~\citet{Planck2014}. Briefly, clusters are detected through their tSZ signature across the six highest frequency \textit{Planck} channels, and selected by imposing that their tSZ signal-to-noise ratio $q>6$, and that they are in the area of the sky left unmasked in the analysis (65\,\% of the sky). The cluster catalogue is publicly available in the \textit{Planck} Legacy Archive (hereafter, PLA)\footnote{\url{http://pla.esac.esa.int/pla/}}.

As a measure of the SZ signal for each cluster in the sample, we follow~\citet{Ade2016} and use the signal-to-noise ratio, $q$. This was measured by the \textit{Planck} Collaboration for all the clusters of the MMF3 cosmology sample through their MMF3 pipeline \citep{Planck2016xxvii}.
%This is the SZ signal proxy used in \citet{Ade2016}.
We similarly use the cluster redshifts obtained
by the Planck Collaboration from ancillary data and follow-up observations, as described in~\citet{Planck2016xxvii}.
%As the measurements of the redshifts of each cluster we also use the values measured by the \textit{Planck} collaboration by follow-up observations and used in \citet{Ade2016}.

In order to calibrate the mass bias, we also incorporate in our analysis CMB lensing mass estimates of all the clusters in the MMF3 cosmology sample with measured redshift. This subsample consists of a total of 433 clusters, and it is the same sample as the one used in the CMB lensing calibration presented in \citet{Ade2016}. In the following, we will refer to it as the \textit{Planck} CMB lensing calibration sample. We directly estimate the masses of all the clusters in this sample from \textit{Planck} 2015 full-mission temperature maps. A detailed account of this process is described in Section \ref{estimation}. For those clusters with redshift information, the \textit{Planck} Collaboration also provide estimates of each cluster's mass, $M_{\text{SZ}}$. These are obtained by combining measurements of the cluster SZ observables $\theta_{500}$, the cluster angular scale, and $Y_{500}$, the SZ flux within a $\theta_{500}$ aperture, which are generally strongly correlated given the resolution of \textit{Planck}, with X-ray-calibrated fiducial scaling relations to break the degeneracy; see~\citet{Planck2016xxvii} for further details. From this conditional estimate of $Y_{500}$, the mass proxy $M_{\text{SZ}}$ can be derived from the same scaling relations. As emphasised in \citet{Planck2016xxvii}, $M_{\text{SZ}}$ should be regarded as the expected \emph{hydrostatic mass} of a cluster, given the assumed scaling relations, cluster redshift and distribution of $Y_{500}$ and $\theta_{500}$ derived from the data. In this work, we only use $M_{\text{SZ}}$ to provide an initial angular scale for the matched filter used to estimate the cluster mass from CMB lensing data (Section~\ref{subsec:matchfilt}), and in Section~\ref{naive} where we compare with the lensing masses to provide a simple (but biased) estimate of the hydrostatic mass bias following~\citet{Ade2016}.

\section{Basics of CMB lensing}\label{formalism}

Massive bodies deflect light due to the effect of their gravity, a phenomenon known as gravitational lensing. CMB photons coming from the last scattering surface are therefore deflected, the observed net effect being a remapping of the CMB fluctuations on the sky by some deflection field $\bmath{\alpha}(\hat{\bmath{n}})$ (see \citealt{Lewis2006} for a general review of CMB lensing).  

Let $X$ be an `unlensed' CMB field, i.e., a CMB field as it would have been observed if there was no lensing, where $X$ can be $T$ (the CMB temperature), $Q$, or $U$ (the two linear polarization Stokes' parameters). Lensing remaps the CMB fields so that the lensed field $\tilde{X}(\hat{\bmath{n}})$ along the line-of-sight direction $\hat{\bmath{n}}$ is the unlensed field at $\hat{\bmath{n}} + \bmath{\alpha}(\hat{\bmath{n}})$, i.e., $\tilde{X}(\hat{\bmath{n}}) = X(\hat{\bmath{n}}+\bmath{\alpha})$. 

The deflection field can be written as $\bmath{\alpha} = \nabla_{\hat{\bmath{n}}} \psi$, where $\nabla_{\hat{\bmath{n}}}$ denotes the angular derivative (covariant derivative on the unit sphere, or, in the flat-sky approximation, partial derivative with respect to the two local angular variables) and where $\psi$ is known as the lensing potential. For a flat universe, the lensing potential can be written as \citep{Lewis2006}
\begin{equation}\label{potentiall}
\psi (\hat{\bmath{n}}) = -2 \int_0^{\chi_{\star}} d\chi \frac{\chi_{\star}-\chi}{\chi_{\star}\chi} \Psi (\chi \hat{\bmath{n}},\eta_0-\chi),
\end{equation}
where $\chi_{\star}$ is the comoving distance to last scattering ($\chi_{\star} \approx 14$\,Gpc), $\eta_0$ is the current conformal time, and $\Psi$ is the Newtonian gravitational potential (or, in a general relativistic framework, the Weyl potential). The lensing potential is therefore a weighted integral of the gravitational potential along the undeflected line of sight.

It is often useful to work with the convergence, $\kappa$, which is given by the two-dimensional Laplacian of the lensing potential, 
\begin{equation}
\kappa (\hat{\bmath{n}}) = - \frac{1}{2} \nabla_{\hat{\bmath{n}}}^2 \psi(\hat{\bmath{n}}).
\end{equation}
Thus, the convergence is a weighted integral of the matter overdensity along the line of sight. Therefore, the integrated matter distribution along the line of sight determines how the CMB photons are lensed, which allows estimation of this integrated matter distribution from CMB observations alone. Several methods to reconstruct the lensing convergence (or, equivalently, the lensing potential) exist, the most computationally simple being based on quadratic estimators \citep{Hu2001,Hu2002}.

Lensing by galaxy clusters produces variations typically of order $10\,\mu$K in the measured CMB temperature \citep{Lewis2006}, which are large enough to be probed by experiments like \textit{Planck} in a statistical way. If a cluster density profile is assumed and the cluster redshift is known, the cluster mass can be estimated by, e.g., fitting an expected cluster convergence profile to a non-parametrically reconstructed convergence, or by fitting the cluster model directly to the CMB maps.

\section{Estimation of the \textit{Planck} cluster masses through CMB lensing}\label{estimation}

In this section, we describe the procedure we follow to estimate the masses of the galaxy clusters in the \textit{Planck} CMB lensing calibration sample using \textit{Planck} data. This is similar to that followed in the CMB lensing cluster mass calibration presented in \citet{Ade2016}, but there are likely minor differences in implementation. (\citealt{Ade2016} do not provide full details of their implementation.)

\subsection{Production of clean temperature maps}

\subsubsection{Cluster fields, masks and apodisation}

We use the six highest frequency (100--857\,GHz) Full-Mission 2015 \textit{Planck} temperature maps, from which we cut square patches of side $256$\,arcmin centred on the positions of the 433 clusters of the \textit{Planck} CMB lensing calibration sample. We will refer to these cutouts as the `cluster frequency fields'. Since the \textit{Planck} maps are in the HEALPix pixelisation \citep{Gorski2005}, we extract the cluster frequency fields as Cartesian projections centred at the cluster positions, setting the number of pixels of each field to be $512 \times 512$, with a pixel size of 0.5\,arcmin. For the cluster central positions we use the Galactic coordinates estimated by the \textit{Planck} collaboration through their cluster detection pipeline, which are publicly available in the PLA. We only use temperature maps, given that, for an experiment with the resolution and noise levels of \textit{Planck}, the $TT$ lensing quadratic estimator offers the best performance in terms of signal-to-noise \citep{Hu2002}. 

We cosine apodise the edges of each cluster frequency field over a width of $12.8$\,arcmin to suppress spectral leakage when taking Fourier transforms. We find, using simulated CMB observations, that with such apodisation applied to a flat-sky, non-periodic CMB realisation with resolution and noise levels similar to \textit{Planck}, we can recover the CMB power spectrum after simple scaling to account for the apodised mask. 
We also mask point sources using the union of the \textit{Planck} 100\,GHz and 143\,GHz 2015 temperature point source masks provided in the PLA, and inpaint the masked regions with the method presented in \citet{Gruetjen2015} to reduce spectral leakage in later stages of our pipeline.

Next, we deconvolve the cluster frequency fields with the corresponding \textit{Planck} isotropic instrumental beam at each field frequency (all of them available in the PLA) and with the HEALPix pixel transfer function. We perform these operations as simple multiplications in harmonic space, which, for the small fields we consider reduces to Fourier space.

\subsubsection{Foreground suppression: constrained internal linear combination}

Since we want to investigate how the primary CMB is deflected by each cluster, we need to suppress all the significant contributions to the observed signal that are on top of the primary, lensed CMB. We also need to reduce the instrument noise as much as possible. The most important foregrounds around a cluster are the thermal and the kinetic SZ effects. They add noise and, more troublingly, bias to the mass measurements (e.g., \citealt{Melin2015}). The tSZ effect has the larger amplitude, in many cases larger than the cluster lensing signal by about an order of magnitude \citep{Lewis2006,Yoo2010}. However, it has a very characteristic frequency dependence, different from the thermal CMB spectrum, which is independent of the cluster temperature in the non-relativistic limit \citep{Rephaeli1995}. In this limit, the SZ spectrum is the same for all galaxy clusters. Thus, multi-frequency CMB observations can be combined in order to estimate and subtract the tSZ signal if the cluster gas is, to a good approximation, non-relativistic. On the other hand, the kinetic SZ signal, a Doppler-shift-induced contribution proportional to the cluster velocity along the line of sight, has the same spectral dependence as the primary CMB, so it cannot be removed with multi-frequency observations. Its amplitude can be of the same order of magnitude as the cluster lensing signal \citep{Seljak2000} and, for asymmetric clusters, may be expected to bias the mass measurement at some level.
However, it has been argued through simulations that the bias is not significant for an experiment with the resolution and sensitivity of \textit{Planck} and for clusters with realistic peculiar velocities \citep{Melin2015}. The kSZ does, however, add additional noise to the mass measurements.

We treat the tSZ effect as a signal to be explicitly removed from our fields and the instrumental noise and all the other foregrounds as noise on top of the primary, lensed CMB signal. In order to suppress the tSZ signal and to reduce the noise of our maps, we perform a constrained internal linear combination (CILC) with the six beam- and pixel-deconvolved cluster frequency fields, as described in \citet{Remazeilles2011}. A CILC is a linear combination of a set of different frequency maps in which the weights are chosen in order to extract a signal with a known spectrum, minimising the variance of the output map with the constraint that another signal with a known spectral signature is completely removed. In this case, the signal to be extracted is the primary, lensed CMB, the signal to be removed is the tSZ signal, and the output is a single, tSZ-free temperature map around each cluster. We shall refer to these maps as `cluster temperature fields'.

In more detail, in a CILC the observations (in our case, the six cluster frequency fields around a given cluster) are modelled as \citep{Remazeilles2011}
\begin{equation}
\bmath{x} (p) = \bmath{a} s(p) + \bmath{b}y(p) + \bmath{n}(p).
\end{equation}
Here, $\bmath{x} (p)$ is a six-dimensional vector containing the values of the six frequency fields at the pixel $p$; $s(p)$ is the CMB map to be extracted, which also includes the kSZ signal as it has the same spectral dependence as the CMB; $y(p)$ is the tSZ contribution, to be removed; and $\bmath{n}(p)$ denotes all the other unmodelled signals plus the instrumental noise. The fixed vectors $\bmath{a}$ and $\bmath{b}$ describe the frequency dependence of the CMB (and kSZ) and tSZ signals, respectively, as integrated across the frequency bands of the experiment.

In units of the CMB temperature, the CMB vector $\bmath{a}=(1,1,1,1,1,1)^T$. In the non-relativistic limit, the tSZ vector $\bmath{b}$ is the same for all clusters and has components for the \textit{Planck} frequency bands that are given in \citet{Planck2016viii}.

Once $\bmath{a}$ and $\bmath{b}$ are known, a weight vector $\bmath{w}$ can be constructed such that $\hat{s} (p) = \bmath{w}^T\bmath{x} (p)$ has unit response to the signal $s(p)$ and zero response to the contaminant $y(p)$. These impose the constraints $\bmath{w}^T\bmath{a} = 1$ and $\bmath{w}^T\bmath{b} = 0$. Further demanding that the effective noise variance be minimised leads to weights~\citep{Remazeilles2011}
\begin{equation}
\bmath{w}^T = \frac{\left(\bmath{b}^T \mathbfss{R}^{-1} \bmath{b} \right) \bmath{a}^T \mathbfss{R}^{-1} - \left( \bmath{a}^T \mathbfss{R}^{-1} \bmath{b} \right) \bmath{b}^T \mathbfss{R}^{-1}} {\left(\bmath{a}^T \mathbfss{R}^{-1} \bmath{a}\right) \left(\bmath{b}^T \mathbfss{R}^{-1} \bmath{b}\right)-\left(\bmath{a}^T \mathbfss{R}^{-1} \bmath{b}\right)^2},
\end{equation}
where $\mathbfss{R}=\langle \bmath{n} \bmath{n}^T\rangle$ is the inter-frequency covariance matrix of the six frequency maps. Here, we assume this is constant across each cluster frequency field and estimate it empirically from the data by summing over pixels.\footnote{Note that adding (symmetric) outer products to $\mathbfss{R}$ involving only $\bmath{a}$ and/or $\bmath{b}$ does not change the weights.}

\subsection{Lensing convergence estimation: quadratic estimators}

The next step is to use our set of cluster temperature fields to estimate the lensing convergence around the clusters in our sample. We perform this estimation with a modified version of the well-known CMB lensing $TT$ quadratic estimator, first described in \citet{Hu2001}, since for an experiment of the resolution and sensitivity of \textit{Planck}, the $TT$ quadratic estimator is very close to the optimal lensing estimator \citep{Raghunathan2017}. 

CMB lensing by a fixed convergence $\kappa(\bmath{x})$ destroys the statistical isotropy of the CMB, coupling Fourier modes with different wavevectors
$\bmath{l}$ in a $\kappa$-dependent way. The lensing quadratic estimators take advantage of this fact in order to estimate $\kappa$. The input of a quadratic estimator is a pair of observed CMB fields defined in the same region of the sky, which can consist of any combination of $T$, $Q$, or $U$; the input pair of the $TT$ estimator is simply two copies of the same temperature map. Its output is an estimate $\hat{\kappa}$ of the convergence, which is unbiased at linear order in $\kappa$.

We use the following form of the $TT$ quadratic estimator in order to estimate $\kappa$ around a given cluster:
\begin{equation}\label{quadratic}
\hat{\kappa} (\bmath{L}) = -N(\bmath{L})  \frac{2}{L^2} \int \frac{d^2 \bmath{x}}{2 \pi} e^{-i \bmath{L} \cdot \bmath{x}} \bmath{\nabla} \cdot \left[F_1(\bmath{x}) \nabla F_2(\bmath{x}) \right],
\end{equation}
where $N(\bmath{L})$ is a normalisation that can be written as
\begin{equation}\label{var}
N(\bmath{L}) = \frac{L^4}{4} \left( \int \frac{d^2 \bmath{l}}{(2 \pi)^2} \frac{\left[(\bmath{L} - \bmath{l}) \cdot \bmath{L} C^{\tilde{T}\tilde{T}}_{|\bmath{l} - \bmath{L}|} + \bmath{l} \cdot \bmath{L} C^{\tilde{T}\tilde{T}}_l \right]^2}{2 C_l^{\hat{T} \hat{T}} C_{|\bmath{l} - \bmath{L}|}^{\hat{T} \hat{T}}} \right)^{-1}.
\end{equation}
Here, $C^{\tilde{T}\tilde{T}}_l$ is the power spectrum of the lensed CMB temperature, and $C^{\hat{T}\hat{T}}_l$ is the (total) power spectrum of the corresponding cluster temperature field. We have also defined two filtered temperature fields,
\begin{equation}\label{f}
F_i(\bmath{l}) \equiv f_i(l) \hat{T}(\bmath{l}),  
\end{equation}
%
% and
% %
% \begin{equation}\label{lg}
% F_2(\bmath{l}) \equiv   f_2(l) \hat{T} (\bmath{l}),
% \end{equation}
for $i=1,2$,
where $\hat{T} (\bmath{l})$ is the corresponding cluster temperature field and the filters are defined as
\begin{equation}
f_1(l) \equiv \frac{1}{C_l^{\hat{T} \hat{T}}},
\end{equation}
and
\begin{equation}
   f_2(l) \equiv 
\begin{cases}
    \frac{C_l^{\tilde{T} \tilde{T}}}{C_l^{\hat{T} \hat{T}}} & \text{if } l \leq l_f, \\
    0              & \text{if } l > l_f. \\
\end{cases}
\end{equation}
Note that $F_1(\bmath{l})$ is the inverse-variance-filtered temperature field, $F_2(\bmath{l})$ is the Wiener-filtered temperature field (for $l\leq l_f$),
and the quadratic estimator~\eqref{quadratic} involves the product of the former with the gradient of the latter.

Our $TT$ quadratic estimator differs from the version originally described in \citet{Hu2001} in two respects. First, we choose $f_2(l)$ to be zero for $l > l_f$. We take $l_f = 2000$. This ensures that the estimator remains unbiased in regions around galaxy clusters, where second and higher orders of $\kappa$ become important and the usual filtering yields biased convergence estimates~\citep{Hu2007}. In addition, we use the lensed CMB power spectrum, $C^{\tilde{T}\tilde{T}}_l$, in $f_2(l)$ and in the 
normalisation, rather than the unlensed CMB power spectrum, as this gives (approximately) the correct response of the estimator $\hat{\kappa}(\bmath{L})$ to lenses at wavevector $\bmath{L}$ averaged over all other lensing modes~\citep{Lewis2011,Hanson2011}.

We estimate $C^{\hat{T}\hat{T}}_l$ empirically for each of our cluster temperature fields to account for the variation of residual foregrounds and instrumental noise across the sky. We calculate $C^{\tilde{T}\tilde{T}}_l$ with CAMB\footnote{\url{http://camb.info/}} assuming our fiducial cosmology. We then extract the central $256 \times 256$ pixels from our cluster temperature fields, and it is on these 128\,arcmin $\times$ 128\,arcmin fields (`reduced cluster temperature fields') that we reconstruct the lensing convergence. The reason for doing this is as follows. We first produce larger cluster temperature fields in order to obtain less noisy local power spectrum estimates. We then choose smaller fields on which to perform the lensing reconstruction to reduce the effect of neighbouring clusters and large-scale structure that is correlated with the cluster, which could contaminate the mass measurement of the central cluster.

Before performing the lensing reconstruction, we cosine apodise each reduced cluster temperature field over a width of $6.4$\,arcmin and then apply the same point source mask used to mask the corresponding set of frequency maps. We inpaint the source-masked regions with constrained Gaussian realisations, as detailed in \citet{Benoit2013}, so that the response of the quadratic estimator is the same as if there was no point-source masking. We account for the apodisation of the edges of the reduced cluster temperature fields by multiplying the reconstructed lensing convergence by a factor $1/\sqrt{f_4}$, where  $f_4$ is given by \citep{Benoit2013}
\begin{equation}
f_4 = \frac{1}{N_{\textnormal{pix}}} \sum_i w_i^4.
\end{equation}
Here, $N_{\textnormal{pix}}$ is the number of pixels in our reduced cluster temperature fields, $w_i$ is the value of the apodisation mask at pixel $i$, and the sum is over all the pixels. Despite this normalisation being appropriate for measuring the power spectrum of statistically-isotropic lensing fields, we verified with simulations that we accurately recover the cluster convergence with this choice.

A mean field has to be subtracted from the lensing reconstruction to account for non-lensing sources of statistical anisotropy, which the quadratic estimator mistakes for lensing thereby biasing the reconstruction (e.g.,~\citealt{Benoit2013,Namikawa2013}).

The sources of the mean field include inhomogeneous noise across the sky, anisotropic instrumental beams, and masks. All of these sources are present, to a certain degree, in our fields: \textit{Planck} noise levels vary across the sky as does the residual foreground power in the CILC; the \textit{Planck} (effective) instrumental beams are not perfectly isotropic; and our apodisation mask breaks isotropy. (The apodisation actually suppresses a larger mean field that would be present if there was no apodisation at all, due to the non-periodicity of the CMB cluster fields.)
In order to subtract the mean field, we define the following modified quadratic estimator:
\begin{equation}
\hat{\kappa}_{\textnormal{mod}}  =   \frac{1}{\sqrt{f_4}} \left( \hat{\kappa} - \hat{\kappa}_{\textnormal{mf}} \right),
\end{equation}
where, $\hat{\kappa}$ is the quadratic estimator presented above and $\hat{\kappa}_{\textnormal{mf}}$ is an estimate of the mean field.

The mean fields will differ across the cluster fields due to variation in the noise, effective beams and residual foreground levels. (We note that the largest contributions to the mean fields are from the masking, but this is approximately the same for all cluster fields, differing only in the point-source masking/inpainting.) We deal with this by estimating local mean fields from the data using reduced cluster temperature fields at random locations on sub-divisions of the sky. In detail, we divide the sky into $20$ regions in bands of ecliptic latitude, since the instrument noise and beam contributions to the mean field are expected to vary significantly with latitude due to the \textit{Planck} scan strategy. We identify each of the $10$ bands falling in the northern-ecliptic hemisphere with the corresponding band located at its mirror location in the southern-ecliptic hemisphere, yielding a total of $10$ distinct bands. We then estimate a mean field within each band and subtract this from the raw convergence for clusters located with the band. To estimate the mean fields, we randomly select $16\,384$ fields with the same size and pixelisation as the cluster frequency fields, but centred on randomly-located points within each band and within the 2015 \textit{Planck} 60\,\% Galactic mask, which was the mask used in the construction of the MMF3 cosmology catalogue. We apply the pipeline described above to the six \textit{Planck} frequency maps for each randomly-centred frequency field and average the resulting estimated convergences (estimated using $\hat{\kappa}$) within each band. We tested with simulated CMB realisations that the number of random fields used to estimate the mean fields is large enough to allow us to ignore statistical errors in the mean fields in the mass estimates.

The estimated $\kappa$ map around each cluster (`cluster reconstructed potential field') is therefore obtained by applying the quadratic estimator $\hat{\kappa}$ to the corresponding apodised and inpainted reduced cluster temperature field, subtracting an estimate of the mean field, and finally renormalising by $1/\sqrt{f_4}$.
At fixed cluster convergence, the variance of this modified estimator is approximately given by
\begin{equation}
\left\langle \Delta \hat{\kappa}_{\textnormal{mod}} \left(\bmath{L}\right) \Delta \hat{\kappa}_{\textnormal{mod}}^{\ast}  \left(\bmath{L}'\right) \right\rangle = \delta^{(2)}  \left(\bmath{L} - \bmath{L}' \right) N_{\kappa}(\bmath{L}),
\end{equation}
where
\begin{equation}\label{variance}
 N_{\kappa}(\bmath{L}) =  C_L^{\kappa \kappa}  +  N^{(0)} (L) + N^{(1)} (L).
\end{equation}
Here, $C_L^{\kappa \kappa}$ is the lensing potential power spectrum (which we obtain from CAMB) and describes the variance from large-scale structure that is uncorrelated with the cluster; $N^{(0)} (L)$, known as the
$N^{(0)}$ bias, describes Gaussian fluctuations of the CMB and noise that mimic the effects of lensing; and $N^{(1)} (L)$, known as the $N^{(1)}$ bias, depends linearly on $C_L^{\kappa \kappa}$ and arises from alternative couplings of the lensed CMB 4-point function with the two copies of the lensing quadratic estimator~\citep{Kesden2003}.
The $N^{(0)}$ bias is approximately equal to the estimator normalisation $N(\bmath{L})$ given in Eq. (\ref{var}), and we calculate the $N^{(1)}$ bias in the flat-sky limit following \citet{Kesden2003}. 

We perform the quadratic lensing reconstruction with the help of \texttt{quicklens}, a freely-available Python CMB lensing package.\footnote{\url{https://github.com/dhanson/quicklens}}

\subsection{Matched filtering}
\label{subsec:matchfilt}

We estimate a cluster mass from each cluster reconstructed convergence field using a matched filter, as presented in~\citet{Melin2015} and used in the \textit{Planck} analysis in~\citet{Ade2016}. The method consists of matched filtering the estimated convergence, $\hat{\kappa}_{\text{mod}}$, with a model for the cluster convergence (the lensing potential is matched filtered in \citealt{Melin2015}, but both procedures are equivalent).

\subsubsection{Cluster model}

Following \citet{Ade2016}, we choose for the cluster mass model a truncated NFW profile \citep{Navarro1997}:
\begin{equation}
   \rho(r) = 
\begin{cases}\label{cluster}
    \frac{\rho_0}{(r/r_s)(1 + r/r_s)^2} & \text{if } r\leq R_{\textnormal{trunc}}, \\
    0              & \text{if } r > R_{\textnormal{trunc}}, \\
\end{cases}
\end{equation}
where $\rho_0$ is a characteristic density; $r_s$ is a characteristic scale radius; and $R_{\textnormal{trunc}}$ is the truncation radius.

In order to be consistent with \citet{Ade2016}, we define the cluster mass in terms of $M_{500}$, the mass contained with a radius $R_{500}$ such that the mean enclosed mass is 500 times that of the critical density at the cluster's redshift, $\rho_c(z)$.
Imposing the definition of $M_{500}$,
\begin{equation}
\rho_0 = \rho_c(z) \frac{500}{3} \frac{c_{500}^3}{\ln(1 + c_{500}) - c_{500}/(1+c_{500})},
\end{equation}
where the concentration parameter is defined by
$c_{500} = R_{500}/r_s $.
As in \citet{Melin2015}, we assume that $c_{500} = 3$, although it is known that it actually varies weakly with cluster mass and redshift \citep{Munoz2011}. 
We further follow \citet{Melin2015} in their choice of truncation radius: $R_{\textnormal{trunc}} = 5 R_{500}$.
Fixing $c_{500}$, for our purposes a cluster is completely specified by two parameters, e.g., $M_{500}$ and redshift $z$.

The lensing convergence of the cluster at sky position $\bmath{\theta}$ from its centre,
$\kappa_{\text{cl}}(\bmath{\theta})$, is related to the (physical) surface density at projected position $\bmath{r}$ from the centre, $\Sigma_{\text{cl}}(\bmath{r})$, through
\begin{equation}
\kappa_{\text{cl}}(\bmath{\theta}) = \Sigma_{\text{cl}}(\bmath{r})/\Sigma_\text{crit}(z),
\end{equation}
where the critical surface density at redshift $z$ is
\begin{equation}
\Sigma_{\text{crit}}(z) = \frac{c^2}{4\pi G} \frac{d_{A,\text{CMB}}}{d_{A,\text{cl}}  d_{A,\text{cl-CMB}}} ,
\end{equation}
which involves the physical angular diameter distances to the CMB last-scattering surface, to the cluster, and between the cluster and the CMB. The projected position at the cluster, $\bmath{r}$, is related to the angular position, $\bmath{\theta}$, through $\bmath{r}=d_{A}(z)\bmath{\theta} = d_{A,\text{cl}}\bmath{\theta}$. The projected surface density can be written as
\begin{equation}
\Sigma_{\text{cl}}(\bmath{r}) = 2 \rho_0 r_s \int_{r/r_s}^{5 c_{500}} \frac{dx}{(1+x)^2\sqrt{x^2-(r/r_s)^2}} ,
\end{equation}
for $|\bmath{r}| \leq R_{\mathrm{trunc}}$.
For a fixed concentration, the integral only depends on $\theta/\theta_s = |\bmath{r}| / r_s$, where the angular scale radius is $\theta_s = r_s / d_A(z)$. It follows that the cluster convergence can be written in the form
\begin{equation}
\kappa_{\text{cl}}(\bmath{\theta}) = \kappa_0 \kappa_{\text{t}}(\bmath{\theta};\theta_s) ,
\end{equation}
where the circularly-symmetric template function $\kappa_{\text{t}}(\bmath{\theta};\theta_s)$ depends only on $\theta/\theta_s$. We choose the normalisation such that $\kappa_{\text{t}}=1$ at the scale radius, so that $\kappa_0$ is the cluster convergence there. It follows that
\begin{equation}
\kappa_0 = \frac{2\rho_0 r_s}{\Sigma_{\text{crit}}(z)} \int_1^{5 c_{500}} \frac{dx}{(1+x)^2\sqrt{x^2-1}} ,
\end{equation}
and so, at fixed concentration,
\begin{equation}
\kappa_0 \theta_s^2 \propto \frac{M_{500}}{\Sigma_{\text{crit}}(z) d_A^2(z)} ,
\end{equation}
i.e., the integrated convergence within an aperture of radius $\theta_s$ is proportional to the cluster mass.

\subsubsection{Matched filter}\label{submatched}

Given an estimate of a cluster angular size $\theta_s$, we filter its reconstructed convergence field with the template $\kappa_{\mathrm{t}}$ to form an estimator for $\kappa_0$:
\begin{equation}\label{filtval}
\hat{\kappa}_0 = \left[\int \frac{d^2 \bmath{L}}{2 \pi} \frac{|\kappa_{\textnormal{t}} (\bmath{L})|^2}{N_{\kappa}(\bmath{L})} \right]^{-1} \int \frac{d^2 \bmath{L}}{2 \pi} \frac{\hat{\kappa}(\bmath{L})  \kappa_{\textnormal{t}}^\ast (\bmath{L}) }{N_{\kappa}(\bmath{L})},
\end{equation}
where $N_{\kappa} (\bmath{L})$ is the variance of $\hat{\kappa}(\bmath{L})$ given in
Eq.~(\ref{variance}). The inverse-variance weighting of $\hat{\kappa}$ ensures this estimator has minimum variance, which can be calculated to be
\begin{equation}\label{filtvar}
\sigma_{\kappa_0}^2 \equiv \left\langle (\hat{\kappa}_0 - \kappa_0)^2 \right\rangle = \frac{1}{2\pi} \left[\int \frac{d^2 \bmath{L}}{2 \pi} \frac{|\kappa_{\textnormal{t}} (\bmath{L})|^2}{N_{\kappa}(\bmath{L})} \right]^{-1}.
\end{equation}

In order to use the filter, an estimate of the cluster angular scale is needed. We use the SZ mass proxy $M_{\mathrm{SZ}}$ of each cluster, as provided in the MMF3 catalogue, as the filter mass, $M^{\mathrm{fid}}$, from which we derive an SZ angular size $\theta_s^{\mathrm{fid}}$:
\begin{equation}\label{massdefinition}
M^{\mathrm{fid}} = 500 \frac{4\pi}{3} \left[d_A(z) c_{500} \theta_s^{\mathrm{fid}}\right]^2 \rho_{c}(z) .
\end{equation}
Since $\kappa_0 \theta_s^2 \propto M_{500}$, we can write an estimator of the cluster's mass as 
\begin{equation}
\hat{M}_{500} = M^{\mathrm{fid}} \left(\frac{\hat{\kappa}_0}{\kappa_0^{\mathrm{fid}}} \right),
\end{equation}
where $\kappa_0^{\mathrm{fid}}$ is the template's convergence at its scale radius $\theta_s^{\mathrm{fid}}$. The standard deviation of this estimator is simply
\begin{equation}\label{mnoise}
\sigma_{M_{500}} = M^{\mathrm{fid}} \left(\frac{\sigma_{\kappa_0}}{\kappa_0^{\mathrm{fid}}}\right) .
\end{equation}
If the filter matches exactly the cluster's true profile, $\hat{M}_{500}$ is an unbiased estimator of the cluster's true mass, $M_{500}$. However, as we now discuss,
$\hat{M}_{500}$ is biased at linear order in any mismatch between the template and the true profile. We can avoid such linear bias by instead working with the signal-to-noise on $\kappa_0$ (or, equivalently, on $M_{500}$), defining the observable
\begin{equation}
\msn \equiv \hat{\kappa}_0 / \sigma_{\kappa_0} = \hat{M}_{500} /  \sigma_{M_{500}}.
\end{equation}

\subsubsection{Biases from template errors}
\label{subsubsec:biasfromtemp}

We consider three possible deviations of the template $\kappa_{\text{t}}$ from the true cluster convergence, which will bias the estimator to some extent: mismatch between the filter fiducial angular scale and the true cluster angular scale, actual mismatch between the filter profile and the true profile, and misplacing of the filter position with respect to the cluster's true position (which we refer to as miscentering).

Let us first consider the effect of mismatch between the filter fiducial angular scale and the true cluster angular scale. Consider a cluster of true size $\theta_s$, but filtered with $\theta_s^{\mathrm{fid}}$. The expected value of the signal-to-noise is
\begin{multline}
\langle \msn \rangle = \sqrt{2\pi}\kappa_0 \left[\int \frac{d^2 \bmath{L}}{2 \pi} \frac{|\kappa_{\textnormal{t}} (\bmath{L};\theta_s^{\mathrm{fid}})|^2}{N_{\kappa}(\bmath{L})} \right]^{-1/2} \\ \times \int \frac{d^2 \bmath{L}}{2 \pi} \frac{\kappa_{\mathrm{t}}(\bmath{L};\theta_s) \kappa_{\textnormal{t}}^\ast (\bmath{L};\theta_s^{\mathrm{fid}})}{N_{\kappa}(\bmath{L})} ,
\label{eq:masssignoise1}
\end{multline}
where $\kappa_0$ is the true value of the convergence at the true scale radius. Recalling that the template convergence $\kappa_{\mathrm{t}}(\bmath{\theta};\theta_s)$ depends only on $\theta/\theta_s$, we can write
\begin{equation}
\kappa_{\mathrm{t}}(\bmath{\theta};\theta_s) = f(\theta/\theta_s) ,
\end{equation}
where the function $f$ is circularly symmetric. The Fourier transform $\kappa_{\mathrm{t}}(\bmath{L};\theta_s)$ is therefore related to the 2D Fourier transform of $f$ through
\begin{equation}\label{massstd}
\kappa_{\mathrm{t}}(\bmath{L};\theta_s) = \theta_s^2 f (\theta_s \bmath{L}) .
\end{equation}
Substituting into Eq.~(\ref{eq:masssignoise1}), we have
\begin{multline}
\langle \msn \rangle = \sqrt{2\pi}\kappa_0 \theta_s^2 \left[\int \frac{d^2 \bmath{L}}{2 \pi} \frac{f^2(\theta_s^{\mathrm{fid}}\bmath{L})}{N_{\kappa}(\bmath{L})} \right]^{-1/2} \\ \times \int \frac{d^2 \bmath{L}}{2 \pi} \frac{f(\theta_s\bmath{L}) f(\theta_s^{\mathrm{fid}}\bmath{L})}{N_{\kappa}(\bmath{L})} ,
\label{eq:masssignoise2}
\end{multline}
The dependence of $\langle \msn \rangle$ on $\theta_s^{\mathrm{fid}}$ can be assessed by writing $\theta_s^{\mathrm{fid}} = \theta_s + \delta\theta_s$ and series expanding in $\delta \theta_s$. For lens reconstruction with \text{Planck} data, the reconstruction noise is large on typical cluster scales, i.e., for $\theta_s |\bmath{L}| \ga 1$, so
$\delta \theta_s |\bmath{L}| \ll 1$ for the modes that dominate the integrals in Eq.~(\ref{eq:masssignoise2}) and the series expansion is expected to converge rapidly.
It is straightforward to show that there is no linear dependence of $\langle \msn \rangle$ on $\delta \theta_s$, so the bias from
using the incorrect cluster angular size in the filtering is only second order in the size error.

We can check numerically that this bias on $\msn$ from adopting an incorrect filter size is indeed small. In particular, we consider a cluster with a mass $M_{500} = 0.5 \times 10^{15} M_{\odot}$ at redshift $z=0.2$ and with our truncated NFW profile as its true convergence profile. We compute $\langle \msn \rangle$ with Eq.~(\ref{eq:masssignoise2}), taking the template to be the true convergence profile but scaled at a number of different angular scales $\theta_s^{\mathrm{fid}}$. For the variance $N_{\kappa} (\bmath{L})$ we use that for a $TT$ quadratic estimator reconstruction for an idealised \textit{Planck}-like experiment with a Gaussian beam of full-width at half-maximum equal to $5\,\text{arcmin}$ and noise level of 45\,$\mu$K\,arcmin. The cluster masses associated with each of the filter angular scales considered, $M^{\mathrm{fid}}$, computed with Eq. (\ref{massdefinition}), and their corresponding $\langle \msn \rangle$ values are shown in Table \ref{filtersize}. It can be seen that value of the filter angular scale $\theta_s^{\mathrm{fid}}$, parametrised in Table \ref{filtersize} in terms of the filter mass proxy $M^{\mathrm{fid}}$, has little impact on $\langle \msn \rangle$ if it is within a reasonable range of the cluster's true angular scale $\theta_s$. Since our filter mass proxies for each cluster are their SZ mass proxies, as provided in the MMF3 catalogue, and given that we expect them to be about $1-b\approx 0.7$ times the cluster true masses, we conclude that the choice of filter angular scale has a negligible impact on our analysis. This would have not been the case if we had chosen to use the filter mass estimates directly, as was done in the CMB lensing calibration in \citet{Ade2016}. Indeed, the third column of Table \ref{filtersize} shows the expected values of the cluster mass estimates at each filter angular scale, $\langle \hat{M}_{500} \rangle$. It can be observed that a significant bias in the mass estimates appears if the filter angular scale $\theta_s^{\mathrm{fid}}$ is different from the truth, $\theta_s$ -- in our test, that corresponding to $M_{500} = 0.5 \times 10^{15} M_{\odot}$. Indeed, $\hat{M}_{500}$ is biased at linear order, while $\msn$ is biased only at second order. We therefore use $\msn$ as our lensing mass observable.

%In order to use the filter, an estimate of the cluster angular size is needed. We use the SZ mass proxy $M_{\mathrm{SZ}}$ of each cluster, as provided in the MMF3 catalogue, to derive an SZ angular size $\theta_s^{\mathrm{fid}}$ from
%
%\begin{equation}\label{massdefinition}
%M_{\mathrm{SZ}} = 500 \frac{4\pi}{3} \left[d_A(z) c_{500} %\theta_s^{\mathrm{fid}}\right]^2 \rho_{c}(z) .
%\end{equation}
%
%We reduce the impact of errors in the assumed cluster sizes by working with the signal-to-noise on $\kappa_0$, defining the observable
%
%\begin{equation}
%\msn \equiv \hat{\kappa}_0 / \sigma_{\kappa_0} .
%\end{equation}
%

\begin{table}
\centering
\caption{Results of our numerical test of the dependence of the expected value of the matched filter output on the filter angular scale $\theta_s^{\mathrm{fid}}$. The cluster considered here has a mass $M_{500} = 0.5 \times 10^{15} M_{\odot}$, is at redshift $z=0.2$, and has our truncated NFW profile as its convergence. The templates with which we filter its convergence have functionally the same profile and are also placed at redshift $z=0.2$, but have a number of different masses $M^{\mathrm{fid}}$, which, at fixed redshift, correspond uniquely to a set of different filter angular scales $\theta_s^{\mathrm{fid}}$. The first column shows the filter masses, $M^{\mathrm{fid}}$, that we have considered; the second column shows the corresponding values of $\langle \msn \rangle$; and the third column, the corresponding values of the expected value of the filter mass estimate, $\langle \hat{M}_{500} \rangle$. It can be seen that $\langle \msn \rangle$ is very insensitive to $M^{\mathrm{fid}}$ (and, therefore, to $\theta_s^{\mathrm{fid}}$) if $M^{\mathrm{fid}}$ is reasonably close to the cluster's true mass, whereas $\langle \hat{M}_{500} \rangle$ has a significant dependence on $M^{\mathrm{fid}}$.}
\label{filtersize}
\begin{tabular}{ccc}
\toprule
$M^{\mathrm{fid}}$                       & $\langle \msn \rangle$ & $\langle \hat{M}_{500} \rangle $ \\
$10^{15} M_{\odot}$ & & $10^{15} M_{\odot}$ \\
\midrule
0.01 & 0.2023  & 0.25 \\
0.10 & 0.2136  & 0.36 \\
0.35 & 0.2164  & 0.46 \\
0.50 & 0.2165  & 0.50 \\
1.00 & 0.2159  & 0.58 \\
10.00 & 0.2055 & 0.98  \\
\bottomrule
\end{tabular}
\end{table}

%\AC{Need to rework this section a bit in light of using $\msn$. In particular, can probably argue generally that mismatch between the template convergence and the truth only biases $\msn$ at second order. For miscentering, the template and truth are related by
%
%\begin{equation}
%\kappa_{\mathrm{t}}^{\mathrm{true}}(\bmath{L}) = \kappa_{\mathrm{t}}(\bmath{L};\theta_s) e^{i \bmath{L} %\cdot \bmath{d}} ,
%\end{equation}
%
%assuming the angular size is correct. Here, $\bmath{d}$ is the centering error. Expect the bias in $\msn$ will be $O(L_\ast^2 d^2)$, where $L_\ast$ is some typical multipole where the reconstruction noise on $\kappa$ starts to grow significantly (and is at most a  few hundred for \textit{Planck}).}

Let us now consider the more general case of mismatch between the filter profile and the mean true cluster profile at given cluster mass and redshift. Following an analogous argument to that for the bias from mismatch in angular size, we argue that profile mismatch biases $\hat{M}_{500}$ to linear order, but $\msn$ only to second order. If the mismatch is small, the bias on $\msn$ will therefore also be small.

It should also be noted that, even if the mean cluster profile at given mass $M_{500}$ and redshift $z$ matches our model, the profile of a given real cluster with mass $M_{500}$ and redshift $z$ can be thought of as a `noisy' realisation of such a mean profile, with the scatter arising from triaxiality and large-scale structure correlated with the cluster. 
This intrinsic scatter is not modelled in our matched filtering process, which only includes variance arising from the lensing reconstruction noise and from uncorrelated large-scale structure, so is still present in our $\msn$ measurements. As we explain in Section~\ref{bayesian}, we do, however, account for intrinsic scatter in the mass measurements in our likelihood.

Finally, let us consider the possibility of miscentering of the filter with respect to the cluster's true position. We centre the filter at the midpoint of each reduced cluster temperature field, corresponding to the SZ-estimated position of the cluster centre as provided in the \textit{Planck} MMF3 catalogue. These positions are, however, not known perfectly. The \textit{Planck} MMF3 catalogue includes an estimate of the 95\,\% confidence interval of the positional uncertainty of each cluster due to the \textit{Planck} beam. Its mean and median values over the cluster sample are both around $2.4$\,arcmin, which implies a typical offset of about $1$\,arcmin. In addition, there is another source of miscentering arising from the offset between the SZ-estimated centre and the centre of mass of each cluster, which is where we ought to place our template. This offset is difficult to estimate, but we expect it to be smaller than the offset due to the beam. Indeed, using hydrodynamical simulations, \citet{Gupta2017} find that for about 80\,\% of the clusters the typical value of this offset is of about $0.04R_{500}$. For a typical cluster in our sample, this translates into an offset of about $0.25$\,arcmin, which, added in quadrature to the offset due to the beam, does not increase it significantly. We can therefore expect a typical miscentering angle, $d$, of about $1$\,arcmin. Assuming that the cluster's mean true profile exactly matches the filter, the filter and the true mean profile are related by 
\begin{equation}
\kappa_{\mathrm{t}}^{\mathrm{true}}(\bmath{L}) = \kappa_{\mathrm{t}}(\bmath{L};\theta_s) e^{i \bmath{L} \cdot \bmath{d}} ,
\end{equation}
where $\bmath{d}$ is the centering error. It is straightforward to show that the bias this induces in $\msn$ is only second order in $d$ (see Appendix~\ref{appendix1} for details). We expect the leading-order bias to be negative, following the $O(\bmath{L}\cdot\bmath{d})^2$ term in the expansion of $\exp(i \bmath{L}\cdot\bmath{d})$, with a value of around $3\,\%$ for $d = 1$\,arcmin in the case in which the filter variance $N_{\kappa} (\bmath{L})$ is appropriate for a $TT$ quadratic estimator reconstruction for a \textit{Planck}-like experiment.
%with a Gaussian beam of $\mathrm{FWHM} = 5$\,arcmin and noise levels of 45\,$\mu$K\,arcmin.
In this scenario, the next-order non-vanishing contribution to the bias is $O(d^4)$ and is at the $10^{-4}$ level, and thus negligible.

We do not explicitly correct for these small, second-order residual biases in our lensing measurements. We do, however, include an effective CMB lensing mass bias parameter in our likelihood analysis, as explained in Section \ref{bayesian}, which we marginalise over to account for the biases.

\section{Pitfalls of simple estimation of the SZ mass bias}\label{naive}

As described in \citet{Ade2016}, the mass bias $1-b$ is introduced in the \textit{Planck} cluster counts analysis in order to account for a possible bias in the X-ray-derived masses used to calibrate the SZ--mass scaling relations. It enters the analysis by multiplying the true cluster mass, $M_{500}$, wherever $M_{500}$ appears in the scaling relations.
%We remind that $M_{500}$ is defined, as usual, as the mass contained within a sphere within which the mean density is 500 times the critical density of the Universe at the cluster's redshift.

As proposed in \citet{Ade2016}, a possible way to estimate the mass bias is to construct the following estimator of $1/(1-b)$:
\begin{equation}
\hat{\frac{1}{1-b}} = \frac{M_{\textnormal{lens}}}{M_{Y_z}},
\label{eq:simplerecipb}
\end{equation}
where $M_{\textnormal{lens}}$ is a lensing estimate of the cluster mass and $M_{Y_z}$ is the corresponding scaling-relation-derived mass. Averaging over an ensemble of clusters, this estimator is an unbiased estimator of $1/(1-b)$ if the following conditions are met: (i) the $M_{\textnormal{lens}}$ measurements are unbiased estimates of $M_{500}$; (ii) the $M_{Y_z}$ measurements are unbiased estimates of the mean SZ mass at a given $M_{500}$, which is what we can define as $(1-b) M_{500}$; and (iii) the scatter on the mean SZ mass at a given $M_{500}$ can be neglected, that is, the $M_{Y_z}$ measurements are precisely $(1-b) M_{500}$. Indeed, if these conditions are satisfied,
\begin{equation}
\left\langle \hat{\frac{1}{1-b}} \right\rangle = \left\langle \frac{M_{\textnormal{lens}}}{M_{Y_z}} \right\rangle = \frac{\left\langle M_{\textnormal{lens}}\right\rangle}{M_{Y_z}} = \frac{M_{500}}{(1-b)M_{500}} = \frac{1}{1-b},
\end{equation}
where angular brackets denote averaging over the cluster sample with some appropriate weighting to minimise the variance. It should be noted that even with such weighting, this estimator is not in general the optimal estimator of the mass bias. 

The conditions needed for this estimator of $1/(1-b)$ to be unbiased are not fully met
for the \textit{Planck} CMB lensing calibration sample, yielding, in general, an incorrect estimate of the mass bias. A detailed quantification of the size of the error, and the extent to which the underlying assumptions are invalid, is beyond the scope of this paper. However, here we offer a brief, qualitative description of the unsuitability of two of the three assumptions that are necessary for the estimator to be accurate. 

First, as already noted in Section \ref{estimation}, it is known that, even in the limit of an unbiased lensing convergence reconstruction around a cluster, the matched filtering process generally gives biased mass estimates due to, e.g., mismatch between the true and template convergence profiles and miscentering. If unaccounted for, as in the calibration presented in \citet{Ade2016}, these biases in the lensing mass estimates propagate directly into $1/(1-b)$.

Secondly, at a given $M_{500}$, the SZ mass estimates in the sample, $M_{Y_z}$, are biased high compared to the true mean SZ mass, $(1-b) M_{500}$. This is due to clusters in our calibration sample being selected through a cutoff on the SZ signal-to-noise, $q$ (plus the additional neglect of six clusters for which there are no redshift measurements available, although this should have very little impact on the modelling of the selection of our sample). This selection effect, which has already been studied in the literature (e.g., \citealt{Mantz2010,Allen2011,Nagarajan2018}), can be understood as follows. Since $M_{Y_z}$ are noisy realisations of the mean SZ mass at a given $M_{500}$, close to the selection cutoff the mean of the SZ masses that get included in the sample for a given $M_{500}$ is necessarily larger than the true population mean $(1-b)M_{500}$. As a consequence, $M_{Y_z}$ becomes a biased estimator of $(1-b) M_{500}$ close to the cutoff for our sample, even if before selection it was unbiased. We therefore expect the simple estimator~\eqref{eq:simplerecipb} to underestimate $1/(1-b)$, i.e., $1-b$ to be biased high.

This selection effect also has some impact on the lensing mass estimates, despite these not intervening in the selection of the sample. This is because the intrinsic scatter in lensing masses is expected to be correlated with that in the SZ masses to a certain extent. Indeed, for a given cluster, both the SZ and the lensing mass estimates are obtained from quantities integrated along the line of sight. Thus, it is reasonable to expect some correlation in their intrinsic scatter, e.g., due to cluster triaxiality. Correlations between SZ and galaxy weak lensing masses have indeed been observed in realistic simulations of galaxy clusters, with a reported intrinsic correlation coefficient as high as $0.8$ \citep{Shirasaki2016}. Similarly, we can expect the scatter in the CMB lensing mass of a cluster to correlate with that in its SZ mass, biasing the mean lensing mass in the sample high if the correlation is positive. Nevertheless, we expect this to be a small effect for current CMB lensing mass estimates since the measurement errors dominate over the intrinsic scatter.

\section{Joint Bayesian analysis of cluster SZ and mass measurements}\label{bayesian}

In this section we develop our Bayesian model to constrain cosmological parameters, notably $\Omega_{\mathrm{m}}$ and $\sigma_8$, and the SZ mass bias (along with several other nuisance parameters) from joint analysis of the SZ data and cluster mass estimates from CMB lensing. The central object is the likelihood, $\mathcal{L}$, giving the probability of the data given the model parameters. The data we use are as follows:
(i) the total number of clusters in the MMF3 cosmology sample, $N$; (ii) the cluster locations on the celestial sphere, $\{\hat{\bmath{n}}_i\}$; and (iii) the cluster redshifts,
$\{z_{\mathrm{obs},i}\}$, as measured by the \textit{Planck} collaboration in follow-up observations, SZ signal-to-noise ratios, $\{q_{\mathrm{obs},i}\}$, as measured by the \textit{Planck} collaboration through their MMF3 pipeline, and CMB lensing signal-to-noise ratios, $\{p_{\mathrm{obs},i}\}$, as measured in this work (see Section \ref{estimation}). We collectively refer to $z_{\mathrm{obs},i}$, $q_{\mathrm{obs},i}$, and $p_{\mathrm{obs},i}$ as the `mass data point' $D_i$ of cluster $i$. Since redshift measurements are not available for six of the clusters in the MMF3 cosmology sample, the mass data point of each such cluster reduces to $q_{\mathrm{obs},i}$. 

The likelihood can be written as the probability density function of $N$, $\underline{\hat{\bmath{n}}}$, and $\underline{D}$, $P\left( N, \underline{\hat{\bmath{n}}}, \underline{D} \right)$, where $\underline{\hat{\bmath{n}}}$ is the vector whose $i$th-component is $\hat{\bmath{n}}_i$, and $\underline{D}$ is the vector whose $i$-component is $D_i$. This probability density function can be decomposed as
\begin{equation}\label{likelihoodfull}
P\left( N, \underline{\hat{\bmath{n}}}, \underline{D} \right) = P(\underline{D} | N, \underline{\hat{\bmath{n}}}) P(\underline{\hat{\bmath{n}}} | N) P(N).
\end{equation}
The first term, $ P(\underline{D} | N, \underline{\hat{\bmath{n}}})$, which we shall refer to as $\mathcal{L}_{1}$, is the probability of obtaining the cluster mass data points $\underline{D}$ given that $N$ clusters have been included in our sample (the MMF3 cosmology sample) and that their sky locations have been found to be $\underline{\hat{\bmath{n}}}$. The dependence on sky location is important, since foregrounds and the \textit{Planck} instrumental noise vary significantly across the sky. The second term, $P(\underline{\hat{\bmath{n}}} | N)$, which we shall denote with $\mathcal{L}_{2}$, is the probability that, given $N$ clusters have been included in our sample, they are located at the sky locations $\underline{\hat{\bmath{n}}}$. Finally, the third term, $P(N)$, which we shall refer to as $\mathcal{L}_{3}$, is the probability of including a total of $N$ clusters in our cluster sample.

Our likelihood is a natural way to extend the SZ counts formalism in order to incorporate the CMB lensing mass measurements to allow for self-calibration of the SZ mass bias. In the rest of this section, we develop the three factors in Eq. (\ref{likelihoodfull}), making clear the parameters on which they depend.  In Appendix \ref{appendix3} we show how, in general, a likelihood like ours is equivalent to a Poisson counts likelihood in $z$--$q_{\rm obs}$--$p_{\rm obs}$ space (in a suitable limit) and how, in particular, our likelihood can be reduced by marginalisation over $p_{\rm obs}$ to a likelihood similar to the \textit{Planck} SZ counts likelihood.

\subsection{$\mathcal{L}_{1}$: Mass data likelihood}\label{clusterdata}

In order to construct $\mathcal{L}_{1} = P(\underline{D} | N, \underline{\hat{\bmath{n}}})$, we assume that each cluster in our sample is statistically independent of the others. This assumption was also made in \cite{Ade2016}, where it was claimed that the impact on the results of the correlations between the different clusters in the sample due to large-scale clustering is weak. As a consequence, we can write $P(\underline{D} | N, \underline{\hat{\bmath{n}}})$ as a product of the probability density functions of the mass data point of each cluster in the sample:
\begin{equation}\label{mult}
P(\underline{D} | N, \underline{\hat{\bmath{n}}}) = \prod_{i=1}^N P (D_i|\inc,  \hat{\bmath{n}}_i).
\end{equation}
Recall that $D_i$ is the mass data point of each cluster, $D_i = (q_{\mathrm{obs},i},p_{\mathrm{obs},i},z_{\mathrm{obs},i})$ for the clusters with known redshift, and $D_i = (q_{\mathrm{obs},i})$ for the clusters with unknown redshift. The condition in $P(\underline{D} | N, \underline{\hat{\bmath{n}}})$ that $N$ clusters are included in the sample is translated into each $P (D_i|\inc,  \hat{\bmath{n}}_i)$ as the condition that each of the clusters is included in the sample. We denote this condition with ``$\inc$'' in Eq.~(\ref{mult}) and throughout. 

We take $P (D_i|\inc,  \hat{\bmath{n}}_i)$ to have the same functional form for all the clusters with known redshift. The same applies for the clusters with unknown redshift. In this case, $P (D_i|\inc,  \hat{\bmath{n}}_i)$ is obtained by marginalising the probability density function of the mass data point of a cluster with known redshift over the corresponding $p_{\mathrm{obs},i}$ and $z_{\mathrm{obs},i}$.

In the following we describe how we construct $P (D_i|\inc,  \hat{\bmath{n}}_i)$ for a cluster with known redshift\footnote{For generality, we develop the likelihood formalism allowing for scatter in the redshift estimates, although in our implementation with the \textit{Planck} clusters we can ignore such scatter for those clusters with redshift information.}
To avoid clutter in the notation, hereafter we will drop the cluster index, $i$. First, we note that
\begin{equation}\label{eqb}
P (D |\inc,  \hat{\bmath{n}}) = \frac{P (\inc | D, \hat{\bmath{n}}) P (D| \hat{\bmath{n}})}{P (\inc | \hat{\bmath{n}})}.
\end{equation}
Given the selection criterion applied in the construction of the MMF3 cosmology sample, $q_{\mathrm{obs}} > 6$, $P (\inc | D, \hat{\bmath{n}})$ is simply a step function at $q_{\mathrm{obs}} = 6$. In order to obtain the other two terms in Eq.~\eqref{eqb}, we adopt a hierarchical model to link each cluster mass data point, $D = (q_{\mathrm{obs}},p_{\mathrm{obs}},z_{\mathrm{obs}})$, to the true cluster mass, $M_{500}$,
and redshift, $z$, and then assume a probability distribution for $M_{500}$ and $z$ (which is what theory can predict).

Our hierarchical model has two layers between $M_{500}$ and $z$ and the mass data point, $D$. First, $q_{\mathrm{obs}}$, $p_{\mathrm{obs}}$, and $z_{\mathrm{obs}}$ are thought of as noisy realisations of their `true' values, $q_{\mathrm{t}}$, $p_{\mathrm{t}}$, and $z_{\mathrm{t}}$, respectively. These are the values of these quantities that would be obtained after averaging over all sources of `observational noise'. We refer to this layer as `observational scatter', and specify what we mean exactly by observational noise below. Second, $q_{\mathrm{t}}$, $p_{\mathrm{t}}$, and $z_{\mathrm{t}}$ are understood as noisy realisations of some mean $\bar{q} \left( M_{500}, z, \hat{\bmath{n}} \right)$, $\bar{p}\left( M_{500}, z, \hat{\bmath{n}} \right)$, and $\bar{z} \left( M_{500}, z, \hat{\bmath{n}} \right)$, respectively, which are specified at given $M_{500}$, $z$, and cluster sky location, $\hat{\bmath{n}}$. We refer to this layer as `intrinsic scatter', and specify its physical sources below.\footnote{More precisely, as explained below, $\ln \bar{q} \left( M_{500}, z, \hat{\bmath{n}} \right)$ and $\ln \bar{p}\left( M_{500}, z, \hat{\bmath{n}} \right)$ are the means of $\ln q_{\mathrm{t}}$ and $\ln p_{\mathrm{t}}$, respectively.}
With this hierarchical model in mind, we can write the probability density function followed by the mass data point of one single cluster as
\begin{align}\label{fulllike}
P (D| \hat{\bmath{n}}) &= P (q_{\mathrm{obs}},p_{\mathrm{obs}},z_{\mathrm{obs}}| \hat{\bmath{n}}) \nonumber \\
&= \int dq_{\mathrm{t}} dp_{\mathrm{t}} dz_{\mathrm{t}} dM_{500} dz \,  \left[P(q_{\mathrm{obs}},p_{\mathrm{obs}},z_{\mathrm{obs}}| q_{\mathrm{t}}, p_{\mathrm{t}}, z_{\mathrm{t}},\hat{\bmath{n}}) \right. \nonumber \\
&\hspace{1.5cm}\left. \times P( q_{\mathrm{t}}, p_{\mathrm{t}}, z_{\mathrm{t}} | M_{500},z,\hat{\bmath{n}}) P(M_{500},z|\hat{\bmath{n}})\right],
\end{align}
where the first factor of the integrand accounts for the observational scatter, the second one accounts for the intrinsic scatter, and the last one is the unconditioned probability density function followed by $M_{500}$ and $z$. 

Since the selection criterion applied in the construction of the MMF3 cosmology sample depends exclusively on the value of $q_{\mathrm{obs}}$, and given the functional form of the intrinsic and observational scatters, which will be described below, the remaining term on the right of Eq.~\eqref{eqb}, $P (\inc | \hat{\bmath{n}})$, can be written using a simplified version of the hierarchical model. Indeed, we can write
\begin{multline}\label{integr}
P (\inc | \hat{\bmath{n}}) = \int dq_{\mathrm{obs}} dq_{\mathrm{t}} dM_{500} dz\, 
\left[ P (\inc | q_{\mathrm{obs}}, \hat{\bmath{n}}) P(q_{\mathrm{obs}}|q_{\mathrm{t}},\hat{\bmath{n}}) \right. \\
\left. \times P(q_{\mathrm{t}} | M_{500}, z, \hat{\bmath{n}}) P(M_{500},z|\hat{\bmath{n}})
\right]
.
\end{multline}
Here, as we will see, the probability distributions governing the intrinsic and observational scatters on the $q$ variables, $P(q_{\mathrm{t}} | M_{500}, z, \hat{\bmath{n}})$ and $P(q_{\mathrm{obs}}|q_{\mathrm{t}},\hat{\bmath{n}})$, respectively, can be readily obtained from those of the full hierarchical model by marginalising over the corresponding $p$ and $z$ variables. Also, as in Eq. (\ref{eqb}), $P (\inc | q_{\mathrm{obs}}, \hat{\bmath{n}})$ is a step function with the step located at $ q_{\mathrm{obs}} = 6$. 

In the following we describe how we compute the mean signal-to-noise at given $M_{500}$ and $z$ for the SZ, CMB lensing, and redshift measurements, and the specific models we adopt for the intrinsic scatter, the observational scatter, and for the unconditioned probability density function of $M_{500}$ and $z$.

\subsubsection{Mean quantities: $\bar{q}$, $\bar{p}$, and $\bar{z}$}

Let us consider a cluster with some given values of $M_{500}$ and $z$, and at sky location $\hat{\bmath{n}}$. As stated above, we assume that these three variables specify some mean values of the SZ signal-to-noise, the CMB lensing mass signal-to-noise, and the redshift of each cluster, $\bar{q} \left( M_{500}, z,\hat{\bmath{n}} \right)$, $\bar{p}\left( M_{500}, z , \hat{\bmath{n}}\right)$, and $\bar{z} \left( M_{500}, z,\hat{\bmath{n}} \right)$.

First, let us consider the mean CMB lensing mass signal-to-noise. As detailed in Section \ref{estimation}, for each cluster we filter the noisy reconstruction of the lensing convergence around its location with a truncated NFW profile in order to obtain an estimate of its lensing mass and the signal-to-noise on this.

In our hierarchical model, we assume that clusters of some $M_{500}$ and $z$ have, on averaging over cluster shape and correlated and uncorrelated large-scale structure, projected mass distributions and so lensing convergences described by the truncated NFW profile specified in Section \ref{estimation} at the true redshift, $z$, but at a biased mass. Matched filtering the reconstructed convergence with an assumed profile  to estimate the mass can introduce further biases due to miscentering and profile mismatch. We assume that the net effect of all sources of bias is to give a mean lensing mass signal-to-noise for clusters of true mass $M_{500}$, redshift $z$, and sky location $\hat{\bmath{n}}$ equal to that for a truncated NFW cluster at redshift $z$ and mass $(1 - b_{\mathrm{CMBlens}})  M_{500}$ filtered at this same (biased) mass scale and redshift:
\begin{equation}\label{mfsnr}
\bar{p}\left( M_{500}, z , \hat{\bmath{n}}\right) = \frac{(1 - b_{\mathrm{CMBlens}})M_{500}}{\sigma_{M_{500}}\left[(1 - b_{\mathrm{CMBlens}})M_{500},z,\hat{\bmath{n}}\right]} .
\end{equation}
Here, $\sigma_{M_{500}}\left[(1 - b_{\mathrm{CMBlens}})M_{500},z\right]$ is the matched filter noise, given by Eq. (\ref{mnoise}). This introduces a dependence of $\mathcal{L}_{1}$ on the sky location of each cluster, since the lensing reconstruction noise varies across the sky. We assume that the lensing mass bias $b_{\mathrm{CMBlens}}$ is constant across the sample. At the resolution and noise levels of \textit{Planck}, $\bar{p}$ scales roughly linearly with $(1-b_{\mathrm{CMBlens}})M_{500}$.

% In our hierarchical model, we assume that a cluster of some $M_{500}$ and $z$ has, on average - averaging over lensing reconstruction noise, cluster shape, and correlated and uncorrelated LSS -, the truncated NFW lensing convergence profile specified in Section \ref{estimation} evaluated at the true redshift, $z$, and at biased mass $(1 - b_{\mathrm{CMBlens}})  M_{500}$. $1-b_{\mathrm{CMBlens}}$, which we assume to be a constant, parametrises this bias. We introduce it in order to account for the small negative bias due to miscentering, not corrected for in previous stages of our pipeline (see Section \ref{estimation}), and from the possible residual bias arising from cluster profile mismatch.

% As shown in Section \ref{estimation}, the mean signal-to-noise of a matched filter measurement is effectively independent from the angular scale of the filter. A cluster with given $M_{500}$ and $z$ and at a sky location $\hat{\bmath{n}}$ can be therefore associated with a unique mean signal-to-noise. Matched filtering this assumed mean profile with the same truncated NFW profile at the same mass, $(1 - b_{\mathrm{CMBlens}})  M_{500}$, and at the same redshift, $z$, gives as mass output ($\langle \hat{M}_{500}\rangle$) the cluster mean lensing mass, $(1 - b_{\mathrm{CMBlens}})  M_{500}$. The mean CMB lensing signal-to-noise at $M_{500}$ and $z$, and at sky location $\hat{\bmath{n}}$ can be thus written as

For the mean SZ signal-to-noise, $\bar{q}$, since we use the measured SZ signal-to-noise ratios $q_{\mathrm{obs}}$ as provided in the \textit{Planck} MMF3 catalogue, we compute $\bar{q}$ at given cluster mass, redshift, and sky location in exactly the same way as the \textit{Planck} team (see \citealt{Planck2014}, \citealt{Ade2016}, and \citealt{Planck2016xxvii}). That is, we write the mean SZ signal-to-noise at $M_{500}$, $z$, and sky location $\hat{\bmath{n}}$ as 
\begin{equation}\label{snr}
\bar{q} (M_{500},z,\hat{\bmath{n}})= \frac{\bar{Y}_{\mathrm{SZ}} \left[(1 - b_{\mathrm{SZ}})M_{500},z\right]}{\sigma_{\mathrm{f}}\left[\theta_{500}((1 - b_{\mathrm{SZ}})M_{500},z),\hat{\bmath{n}}\right]}.
\end{equation}
Here, $\sigma_{\mathrm{f}}(\theta_{500},\hat{\bmath{n}})$ is the noise of the multifrequency matched filter used to detect the clusters evaluated at the SZ angular scale of the cluster, $\theta_{500}$, where
\begin{equation}\label{thetaeq}
\theta_{500} = \theta_{\star} \left( \frac{h}{0.7} \right)^{-2/3} \left( \frac{(1 - b_{\mathrm{SZ}}) M_{500}}{3 \times 10^{14} M_{\odot}} \right)^{1/3} E^{-2/3} (z) \left( \frac{d_A (z)}{500 \mathrm{Mpc}} \right)^{-1} .
\end{equation}
The constant $\theta_{\star} = 6.997$\,arcmin, $E(z) = H(z)/H_0$, where $H(z)$ is the Hubble parameter and the Hubble constant $H_0 = 100 h \, \mathrm{km}\,\mathrm{s}^{-1}\,\mathrm{Mpc}^{-1}$, and $d_A (z)$ is the angular diameter distance to redshift $z$. The SZ mass bias parameter $b_{\mathrm{SZ}}$ is introduced to account for any differences between the X-ray mass estimates that are used in the calibration of the SZ scaling relations and the true masses, as discussed in Section~\ref{sec:intro}.
The mean integrated Comptonisation parameter in Eq.~\eqref{snr}, $\bar{Y}_{\mathrm{SZ}}$, is given by the scaling relation \citep{Ade2016}
\begin{equation}
E^{-\beta}(z) \left( \frac{d_A^2(z) \bar{Y}_{\mathrm{SZ}}}{10^{-4}\, \mathrm{Mpc}^2} \right) = Y_{\star} \left( \frac{h}{0.7} \right)^{-2+\alpha} \left(\frac{(1 - b_{\mathrm{SZ}}) M_{500}}{6\times 10^{14} M_{\odot}} \right)^{\alpha},
\end{equation}
where $Y_{\star}$, $\alpha$, and $\beta$ are parameters that need to be calibrated. We compute $\sigma_{\mathrm{f}}(\theta_{500},\hat{\bmath{n}})$ by interpolating over the tabulated values used in the likelihood analyses in \citet{Ade2016}; their likelihood code is freely available as part of the COSMOMC package.\footnote{\url{https://cosmologist.info/cosmomc/}}
As for the CMB lensing case, $\sigma_{\mathrm{f}}(\theta_{500},\hat{\bmath{n}})$ introduces an additional dependence of $\mathcal{L}_{1}$ on the sky location of each cluster.

%Let us now consider the mean SZ signal-to-noise. We model the SZ signal-to-noise measurements in exactly the same way as in \citet{Ade2016}. In such work, like in this one, the mean SZ signal-to-noise at given $M_{500}$, $z$, and sky location $\hat{\bmath{n}}$ is obtained analogously to the mean CMB lensing signal-to-noise, since the SZ signal-to-noise measurements were obtained by the \textit{Planck} collaboration through their MMF3 pipeline, which also makes use of a matched filter, this time a multi-frequency matched filter across the six highest frequency maps of \textit{Planck}. The matched filter used is the comptonisation profile from \citet{Arnaud2010}. In analogy with the CMB lensing case, clusters are assumed to follow, on average, such profile, with the mass coming into the profile also being biased with respect to $M_{500}$. It is taken to be $(1 - b_{\mathrm{SZ}}) M_{500}$, where $(1 - b_{\mathrm{SZ}})$, parametrising the SZ mass bias, is also assumed to be a constant. This bias is denoted as $(1-b)$ in \citet{Ade2016}. 

Finally, as in \citet{Ade2016}, we take the mean redshift at $M_{500}$, $z$, and $\hat{\bmath{n}}$ to be simply $z$, i.e.,
\begin{equation}
\bar{z} \left( M_{500}, z ,\hat{\bmath{n}}\right)=z .
\end{equation}

\subsubsection{Intrinsic scatter}\label{intrscatter}

The first layer in our hierarchical model is what we refer to as intrinsic scatter. For the SZ and CMB lensing measurements, by intrinsic scatter we mean all sources of statistical scatter on the signal-to-noise measurements that are not incorporated in the noise budget of the matched filtering process. For the CMB lensing measurements, these include deviations of the cluster convergence profile from its assumed mean profile at $M_{500}$ and $z$, deviations which often exhibit a triaxial nature, and contributions to the observed convergence profile from correlated large-scale structure.
For the SZ measurements, intrinsic scatter includes deviations from the mean Comptonisation profile, also often of a triaxial nature. Finally, as in \citet{Ade2016}, we assume that there is no intrinsic scatter on the redshift measurements.

In \citet{Ade2016} a log-normal model for the intrinsic scatter of the SZ signal-to-noise was adopted. Intrinsic scatter on lensing-derived cluster masses has also been shown to be approximately log-normal (see, e.g., \citealt{Becker2011}), and the SZ and CMB lensing measurements are also expected to be intrinsically correlated, since they are both obtained from quantities integrated along the line of sight (see, e.g., \citealt{Shirasaki2016}). We therefore adopt the following probability distribution for the intrinsic scatter:
\begin{equation}\label{eqintr}
 P( q_{\mathrm{t}}, p_{\mathrm{t}}, z_{\mathrm{t}} | M_{500},z,\hat{\bmath{n}}) =  \frac{1}{q_{\mathrm{t}}p_{\mathrm{t}}} P( \ln q_{\mathrm{t}}, \ln p_{\mathrm{t}} | M_{500},z,\hat{\bmath{n}}) \delta (z_{\mathrm{t}} - z),
\end{equation}
where the delta function reflects the absence of intrinsic scatter on the redshift measurements, and where $P( \ln q_{\mathrm{t}}, \ln p_{\mathrm{t}} | M_{500},z,\hat{\bmath{n}})$ is a bivariate Gaussian with means $\ln \bar{q} (M_{500},z,\hat{\bmath{n}})$ and $\ln \bar{p} (M_{500},z,\hat{\bmath{n}})$, respectively, and a covariance matrix that we parametrise with its two associated standard deviations, $\sigma_{\mathrm{SZ}}$ and $\sigma_{\mathrm{CMBlens}}$, respectively, and a correlation coefficient, $r_{\mathrm{CMBlens-SZ}}$. We take $\sigma_{\mathrm{SZ}}$, $\sigma_{\mathrm{CMBlens}}$, and $r_{\mathrm{CMBlens-SZ}}$ to be constants.
%Here and throughout, $\ln$ denotes the natural logarithm.

\subsubsection{Observational scatter}\label{obsscatter}

The second layer in our hierarchical model is what we call observational scatter. For the CMB lensing and SZ measurements, it accounts for the scatter caused by what is treated as observational noise in the matched filtering processes that yield the values of $q_{\mathrm{obs}}$ and $p_{\mathrm{obs}}$ of each cluster. For example, for the CMB lensing masses, observational scatter arises from the reconstruction noise and large-scale structure that is uncorrelated with the cluster. We assume that the observational scatter is Gaussian for both SZ and lensing measurements. This assumption was also made for the SZ measurements in \citet{Ade2016}, and should be also a good approximation for the CMB lensing measurements, since they are obtained by matched filtering a reconstructed convergence map, which involves a sum over many reconstructed modes. We also assume that there is no correlation between the observational scatter of the CMB lensing measurements and that of the SZ measurements. Finally, as in \citet{Ade2016}, we assume that there is no observational scatter on $z_{\mathrm{t}}$. 

We can therefore write the probability distribution function accounting for the observational scatter as
\begin{equation}\label{eqobs}
P(q_{\mathrm{obs}},p_{\mathrm{obs}},z_{\mathrm{obs}}| q_{\mathrm{t}}, p_{\mathrm{t}}, z_{\mathrm{t}},\hat{\bmath{n}}) = P(q_{\mathrm{obs}} | q_{\mathrm{t}}) P(p_{\mathrm{obs}} | p_{\mathrm{t}}) \delta (z_{\mathrm{obs}} - z_{\mathrm{t}}),
\end{equation}
where $P(q_{\mathrm{obs}} | q_{\mathrm{t}})$ and $P(p_{\mathrm{obs}} | p_{\mathrm{t}})$ are Gaussians with means $q_{\mathrm{t}}$ and $p_{\mathrm{t}}$, respectively, and unit standard deviations. The delta function in Eq.~\eqref{eqobs} accounts for the absence of observational scatter on the redshift measurements.

\subsubsection{Unconditioned distribution of $M_{500}$ and $z$}\label{priorm}

The remaining factor of the integrand in Eq.~(\ref{fulllike}) is $P(M_{500},z|\hat{\bmath{n}})$. This is the probability density function followed by $M_{500}$ and $z$ with no conditions imposed, other than the sky location $\hat{\bmath{n}}$. It does not depend on $\hat{\bmath{n}}$ and is simply proportional to the halo mass function, $d^2N/(dV dM_{500})$, times the volume element, $dV/dz$:
\begin{equation}\label{massfunc}
P(M_{500},z|\hat{\bmath{n}}) \propto \frac{d^2N}{dV dM_{500}} (M_{500},z) \frac{dV}{dz} (z).
\end{equation}
As in \citet{Ade2016}, we use the halo mass function from \citet{Tinker2008}. 

\subsubsection{Probability of inclusion, or renormalisation}

Finally, let us revisit the probability for a cluster to be included in the sample, $P (\inc | \hat{\bmath{n}})$, as written down in Eq. (\ref{integr}). From Sections \ref{intrscatter} and \ref{obsscatter}, it should now be apparent that such a decomposition is possible, and that $P(q_{\mathrm{obs}} | q_{\mathrm{t}},\hat{\bmath{n}})$ is just $P(q_{\mathrm{obs}} | q_{\mathrm{t}})$ in Eq. (\ref{eqobs}), and $P(q_{\mathrm{t}} | M_{500}, z, \hat{\bmath{n}})$ is just $P( q_{\mathrm{t}}, p_{\mathrm{t}}, z_{\mathrm{t}} | M_{500},z,\hat{\bmath{n}})$ in Eq. (\ref{eqintr}), marginalised over $p_{\mathrm{t}}$ and $z_{\mathrm{t}}$.

\subsection{$\mathcal{L}_{2}$: Sky location likelihood}
\label{subsec:skylocL}

The second factor of our total likelihood, $\mathcal{L}_{2}$ (see Eq.~\ref{likelihoodfull}), is $P(\underline{\hat{\bmath{n}}}|N)$, the probability that, given $N$ clusters have been included in our sample (the MMF3 cosmology sample), their sky locations are $\underline{\hat{\bmath{n}}}$. Since we assume that clusters are statistically independent from each other, it is simply given by 
\begin{equation}\label{locations}
P(\underline{\hat{\bmath{n}}}|N) = N! \prod_{i=1}^{N} \frac{1}{\bar{N}}\frac{d\bar{N}}{d\Omega} (\hat{\bmath{n}}_i) ,
\end{equation}
where $d\bar{N}/d\Omega$ is the mean number of clusters in the sample per solid angle, and $\bar{N}$ is the mean total number of clusters in the sample, i.e.,
\begin{equation}\label{meann}
\bar{N} = \int d\hat{\bmath{n}} \frac{d\bar{N}}{d\Omega} (\hat{\bmath{n}}),
\end{equation}
where the integration is performed across the region of the sky left unmasked in the construction of the MMF3 cosmology sample. We include the $N!$ factor in Eq. (\ref{locations}) to reflect the fact that the ordering of the elements in $\underline{\hat{\bmath{n}}}$ does not matter.\footnote{The integration measure over the data for $N$ clusters is then $(1/N!)\prod_{i=1}^N d D_i d \hat{\bmath{n}}_i$.}
The mean number of clusters in the sample per solid angle
can be written as
\begin{equation}\label{npersa}
\frac{d\bar{N}}{d\Omega} (\hat{\bmath{n}}) = \int dq_{\textnormal{obs}} dM_{500}  dz\, P \left( q_{\textnormal{obs}} | M_{500},z,\hat{\bmath{n}} \right)  \frac{d^2N}{dV dM_{500}} \frac{d^2V}{dz d\Omega}.
\end{equation}
Here, $q_{\textnormal{obs}}$ is integrated from $6$ to infinity (to meet the inclusion selection),
%and $M_{500}$ and $z$, both from $0$ to infinity,
$d^2N/(dV dM_{500})$ is the halo mass function, and $d^2V/(dz d\Omega)$ is the volume element (see Section \ref{priorm}). The probability density
\begin{equation}
P \left( q_{\textnormal{obs}} | M_{500},z,\hat{\bmath{n}} \right) = \int d q_{\text{t}} \,
P\left(q_{\textnormal{obs}} | q_{\text{t}}\right) P\left(q_{\text{t}}|M_{500},z,\hat{\bmath{n}} \right) ,
\end{equation}
which involves the (marginalised) probability density functions for the observational and intrinsic scatter introduced in Section~\ref{clusterdata}.

\subsection{$\mathcal{L}_{3}$: Poisson likelihood for the total number of clusters}

Finally, the third factor in our total likelihood, $\mathcal{L}_{3}$ (see Eq.~\ref{likelihoodfull}), is $P(N)$, the probability of including a total of $N$ clusters in our sample (the MMF3 cosmology sample). Since we assume that clusters are statistically independent from each other, $P(N)$ is simply a Poisson distribution with expected value $\bar{N}$, where $\bar{N}$ is given by Eq. (\ref{meann}).

\subsection{Model parameters and priors}
\label{subsec:paramsandpriors}

\subsubsection{Cluster parameters} 

For convenience, we summarise below the set of cluster model parameters on which our likelihood, $\mathcal{L}$, depends:
\begin{equation}
\bmath{p}_{\text{cl}} = \left\lbrace 1-b_{\textnormal{SZ}}, 1-b_{\textnormal{CMBlens}} , \sigma_{\textnormal{SZ}},  \sigma_{\textnormal{CMBlens}}, r_{\mathrm{CMBlens-SZ}}, Y_{\star}, \alpha, \beta \right\rbrace.
\end{equation}
A list of these parameters can also be found in Table \ref{table_parameters}, along with their definitions.

\begin{table}
\centering
\caption{Summary of the cluster parameters of our model.}
\label{table_parameters}
\begin{tabular}{@{}ll@{}}
\toprule
Parameter                     & Definition                                                    \\ \midrule
$1-b_{\textnormal{SZ}}$         & SZ mass bias                                                      \\
$1-b_{\textnormal{CMBlens}}$        & CMB lensing mass bias                                             \\
$\sigma_{\textnormal{SZ}}$    & SZ intrinsic scatter                               \\
$\sigma_{\textnormal{CMBlens}}$   & CMB lensing intrinsic scatter                      \\
$r_{\textnormal{CMBlens-SZ}}$ & SZ--CMB lensing intrinsic correlation coefficient     \\
$Y_{\star}$             & Scaling relation normalisation                                   \\
$\alpha$                      & Scaling relation mass exponent                                   \\
$\beta$                       & Scaling relation $E(z)$ exponent                                   \\ \bottomrule
\end{tabular}
\end{table}

We adopt priors on these cluster parameters in our likelihood analysis as follows.
For the SZ parameters, we fix $\beta = 0.66$ (i.e., self-similar evolution), following the main analysis of~\citet{Ade2016}. Although the redshift evolution of the SZ scaling relation is not well constrained observationally, \citet{Ade2016} showed that the main impact of allowing $\beta$ to vary within a broad Gaussian prior, $\beta = 0.66 \pm 0.50$, was to broaden constraints along the $\Omega_{\rm m}$--$\sigma_8$ degeneracy direction, rather than in the perpendicular direction that responds to the absolute cluster mass scale (see their Appendix A.3). We expect similar effects in our analysis, with only a small impact on constraints on the mass bias parameter $1-b_{\text{SZ}}$ of primary interest. We also fix the intrinsic scatter in the SZ signal-to-noise to $\sigma_{\mathrm{SZ}} = 0.173$, the central value of the empirically-derived prior adopted in~\citet{Ade2016}, for reasons of computational efficiency. \footnote{{
The constraints on this parameter from~\citet{Ade2016} for the CCCP and WtG calibrations are prior-driven, the central values being a fraction of a $\sigma$ away from the prior central value. In addition, we find from their corresponding MCMC chains that $\sigma_{\mathrm{SZ}}$ has small correlation with our main parameter of interest, $\sigma_8 \left(\Omega_{\mathrm{m}}/0.33\right)^{0.25}$, with a correlation coefficient of about $0.2$ for both calibrations (see Section~\ref{results} for the motivation behind this parametrisation). We therefore conclude that letting this parameter vary and imposing on it the same prior as in \citet{Ade2016} would have little impact on our results.}} For the remaining parameters in the SZ scaling relation, $\alpha$ and $\log Y_\star$, we adopt the Gaussian priors reported in Table \ref{priorstable}, following~\citet{Ade2016}. Finally, for the SZ mass bias we adopt a flat positivity prior $1- b_{\textnormal{SZ}} \geq 0$. 

For the CMB lensing mass parameters, we impose Gaussian priors on 
$1- b_{\textnormal{CMBlens}}$, $\sigma_{\textnormal{CMBlens}}$, and $r_{\mathrm{CMBlens-SZ}}$
with means and standard deviations given in Table~\ref{priorstable}. In Zubeldia \& Challinor (in prep.), we study the intrinsic bias and scatter of our CMB lensing observable using $N$-body simulations (as was done, e.g., in \citet{Becker2011} for galaxy weak lensing). Our prior on $1- b_{\textnormal{CMBlens}}$ is taken to be consistent with the values we find in Zubeldia \& Challinor (in prep.) in addition to the bias expected from miscentering. Similarly, our prior on $\sigma_{\textnormal{CMBlens}}$ is taken to be consistent with the values we report in Zubeldia \& Challinor (in prep.). Finally, our prior for the correlation between the intrinsic scatter on SZ and CMB lensing mass signal-to-noise ratios, $r_{\mathrm{CMBlens-SZ}}$, is taken to be consistent with the values of the SZ--lensing mass intrinsic correlation reported in~\citet{Shirasaki2016}.

\begin{table}
\centering
\caption{Means and standard deviations of the Gaussian priors adopted in our likelihood analysis of the real cluster data. Parameters not listed have broad flat priors.}
\label{priorstable}
\begin{tabular}{@{}lll@{}}
\toprule
Parameter                       & Mean    & Standard deviation \\ \midrule
$1 - b_{\textnormal{CMBlens}}$  & 0.93    & 0.05               \\
$\sigma_{\textnormal{CMBlens}}$ & 0.20    & 0.05               \\
$r_{\textnormal{CMBlens-SZ}}$   & 0.83    & 0.10               \\
$\alpha$          & 1.79    & 0.08               \\
$\log{Y_{_\star}}$          & -0.19    & 0.02               \\
$100 \theta_{\textnormal{MC}}$  & 1.04093 & 0.00030            \\ \bottomrule
\end{tabular}
\end{table}

\subsubsection{Cosmological parameters}

We also stress the dependence of our likelihood on cosmological parameters, so we can jointly constrain cosmology and cluster physics.
The strongest dependences are on $\Omega_{\mathrm{m}}$, $\sigma_8$, and $H_0$ through the halo mass function, the volume element, and the mean values of the SZ and CMB lensing signal-to-noise ratios at given $M_{500}$, $z$, and sky location. There are only weak dependencies on the baryon density, $\Omega_{\mathrm{b}} h^2$, and spectral index of the primordial curvature perturbations, $n_{\mathrm{s}}$, through the shape of the matter power spectrum around cluster scales that enters the halo mass function. In our analysis, we fix these two parameters to the values $\Omega_{\mathrm{b}} h^2 = 0.0223$ and $n_{\mathrm{s}} = 0.9667$ determined by \textit{Planck}~\citep{Planck2016xiii}. We also fix $\Omega_{K}=0$ and the summed neutrino mass to the minimal value, $\sum m_\nu = 0.06\,\text{eV}$.

We impose broad, flat priors on $\Omega_{\mathrm{m}}$, $\sigma_8$, and $h$.
% $0.1 < \Omega_{\mathrm{m}} < 0.9$, $0.3 < \sigma_8 < 1$, and $0.2 < h < 1$
However, since $h$ is poorly constrained by the cluster data, we further impose a Gaussian prior on the CMB acoustic scale parameter, $\theta_{\text{MC}}$ (see Table~\ref{priorstable}). This parameter is an analytic approximation to the ratio of the sound horizon at CMB last scattering to the angular diameter distance -- see~\citet{Kosowsky2002}, where $\theta_{\text{MC}}$ is their $\mathcal{A}$ -- which at fixed baryon density depends only on $\Omega_{\mathrm{m}}$ and $H_0$ in flat $\Lambda$CDM models. The acoustic scale parameter is very well measured by \textit{Planck} from the acoustic peak locations, and is almost model independent. Our prior is from the analysis reported in~\citet{Planck2016xiii}. We stress that this prior is \emph{geometric} and uses no information on the amplitude of the fluctuations from the CMB power spectrum.

\section{Likelihood validation}\label{validation}

We test the implementation of our likelihood by analysing a set of simulated data, $\left\lbrace N, \underline{\hat{\bmath{n}}}, \underline{D} \right\rbrace$. We produce these data following our model assumptions at fixed values of cosmological and cluster model parameters, and then explore the corresponding posterior distribution with Markov-chain Monte Carlo (MCMC).

\subsection{Simulated data}\label{sd}

In order to generate our set of simulated data, we assume a flat $\Lambda$CDM cosmology with $\Omega_{\mathrm{m}} = 0.315$, $\sigma_8 = 0.811$, $h = 0.674$, and $\Sigma m_{\nu} = 0.06$\,eV. 
We set the cluster model parameters close to the mean values of the Gaussian priors in Table~\ref{priorstable} along with $\beta = 0.66$, $\sigma_{\textnormal{SZ}} = 0.173$ (as fixed by our priors), and $1-b_{\textnormal{SZ}} = 0.62$.
These fiducial values are listed in the final column of Table~\ref{1dconstraints-sim}.

The first step in producing our simulated dataset is to obtain $N$, the total number of clusters that will be in our sample. We obtain it by drawing one sample from a Poisson distribution with mean value $\bar{N}$ given by Eq. (\ref{meann}), evaluated at our assumed values of cosmological and model parameters. We obtain a total of $N=416$ clusters.

We then generate the set of sky locations of these $N$ clusters, $\underline{\hat{\bmath{n}}}$. To do this, in order to produce a catalogue that is statistically as close as possible to the MMF3 catalogue we use the SZ matched filter noise estimates across the sky produced by the \textit{Planck} collaboration and that were used in the construction of the MMF3 catalogue. In these, the SZ matched filter noise is estimated for a set of filter angular scales in a set of patches that cover the whole sky. For each patch, interpolating over these tabulated values and using Eq. (\ref{meann}), where we restrict the integration over the sky to the extent of the patch, we compute the expected number of clusters falling within the patch. For patch $i$, this is $\bar{N}_i$, and the probability for a detected cluster to be in that patch is simply $P_i = \bar{N}_i / \bar{N}$. We then assign a number of clusters to each patch by obtaining $N$ samples from a multinomial distribution with probabilities $P_i$.

Finally, we produce a mass data point, $D_i$, for each of the $N$clusters. We generate each of them in the following way. First, we obtain values of $M_{500}$ and $z$ by drawing a sample from $P(M_{500},z|\hat{\bmath{n}})$ (see Eq.~\ref{massfunc}) using rejection sampling. We then produce values of $q_{\textnormal{t}}$, $p_{\textnormal{t}}$, and $z_{\textnormal{t}}$ by drawing one sample from Eq. (\ref{eqintr}), with the values of $M_{500}$ and $z$ obtained in the previous step and $\hat{\bmath{n}}$ as conditioning information. In this step, for each cluster, $\bar{q}$ is computed with the SZ matched filter noise estimate of the sky patch in which the cluster is located. On the other hand, $\bar{p}$ is computed with the empirical average CMB lensing matched filter noise of all the clusters in the MMF3 cosmology sample falling within the cluster's sky patch. This is for the sake of simplicity, since in our CMB lensing mass pipeline we have produced such noise estimates only at the MMF3 cluster locations. To avoid inconsistencies in the analysis of the \emph{simulated} data, we use these averaged CMB lensing noise estimates in the likelihood in this case (but not in the analysis of the real cluster sample). Finally, we generate values of $q_{\textnormal{obs}}$, $p_{\textnormal{obs}}$, and $z_{\textnormal{obs}}$ by drawing one sample from Eq. (\ref{eqobs}), with the values of $q_{\textnormal{t}}$, $p_{\textnormal{t}}$, and $z_{\textnormal{t}}$ obtained in the previous step as conditioning information. If $q_{\textnormal{obs}} > 6$, we set $(q_{\textnormal{obs}}, p_{\textnormal{obs}}, z_{\textnormal{obs}})$ as the mass data point corresponding to the cluster at sky location $\hat{\bmath{n}}$. If not, we repeat the process again until $q_{\textnormal{obs}} > 6$.

\subsection{Parameter constraints}

In order to validate the implementation of our likelihood, $\mathcal{L}$, we apply it to the simulated dataset, described in Section \ref{sd}, and explore the corresponding posterior distribution with an MCMC. We allow the same set of cosmological and cluster model parameters to vary as in our analysis of the real data,
$\bmath{p} = \left\lbrace \Omega_{\mathrm{m}}, \sigma_8, h, b_{\textnormal{SZ}}, b_{\textnormal{CMBlens}} , \sigma_{\textnormal{CMBlens}}, r_{\mathrm{CMBlens-SZ}} , \alpha, Y_{\star} \right\rbrace$, and impose similar priors on them. For those parameters on which we impose Gaussian priors when analysing the real data (see Table~\ref{priorstable}), we retain the widths of the priors but, where necessary, recentre the means on the input values used to generate the simulations. As for the real data, we impose broad, flat priors on $\Omega_{\mathrm{m}}$, $\sigma_8$, $h$, and $1-b_{\textnormal{SZ}}$, while $\sigma_{\textnormal{SZ}}$ and $\beta$ are fixed to their input values.

\begin{figure}
\centering
\includegraphics[width=0.5\textwidth]{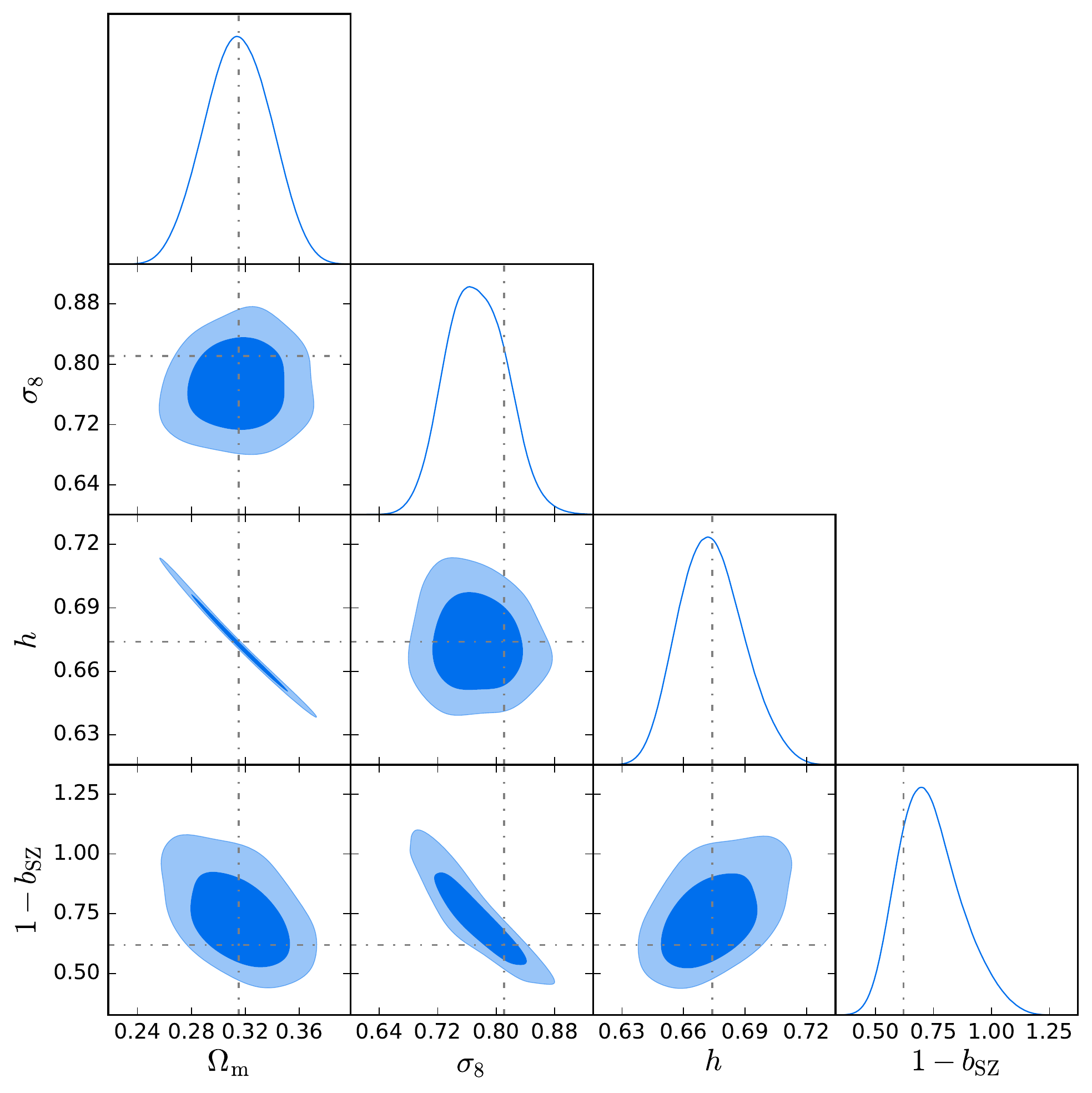}
\caption{Constraints (68\,\% and 95\,\% confidence regions)
on $\Omega_{\mathrm{m}}$, $\sigma_8$, $h$, and $1-b_{\textnormal{SZ}}$ for our simulated cluster sample. One-dimensional, marginalised posterior distributions are also shown. The input values of the parameters are shown as dash-dotted lines.}
\label{corner_sim}
\end{figure}

We explore our posterior using the \texttt{emcee} package, which performs affine-invariant MCMC sampling.\footnote{\url{http://dfm.io/emcee/current/}} The two-dimensional marginalised constraints that we obtain for $\Omega_{\mathrm{m}}$, $\sigma_8$, $h$, and $1-b_{\textnormal{SZ}}$ are shown in Fig.~\ref{corner_sim}, and the one-dimensional marginalised constraints for all the parameters varied in the analysis are given in Table \ref{1dconstraints-sim}. It can be seen that these constraints are consistent with the input parameter values used to construct the simulated data.
The two-dimensional marginalised constraints on the remaining parameters are not shown, but are similarly consistent with their true values. This validates our likelihood implementation and, as long as our Bayesian model remains a good description of the real data, our analysis. Of course, as our validation is limited to one simulation, we cannot test for systematic biases at a level below the statistical errors expected from our real data.

% \begin{table}
% \centering
% \caption{Parameters on which a non-flat prior is imposed in our MCMC analysis of the likelihood of our simulated data. All of them are Gaussian priors; the values of their respective means and standard deviations are specified. }
% \label{priorstablesim}
% \begin{tabular}{@{}lll@{}}
% \toprule
% Parameter                       & Mean    & Standard deviation \\ \midrule
% $1 - b_{\textnormal{CMBlens}}$  & 0.92    & 0.05               \\
% $\sigma_{\textnormal{CMBlens}}$ & 0.23    & 0.05               \\
% $r_{\textnormal{CMBlens-SZ}}$   & 0.77    & 0.10               \\
% $\alpha$          & 1.79    & 0.08               \\
% $\ln{Y_{_\star}}$          & -0.19    & 0.02               \\
% $100 \theta_{\textnormal{MC}}$  & 1.04093 & 0.00030            \\ \bottomrule
% \end{tabular}
% \end{table}

\begin{table}
\centering
\caption{Marginalised constraints (mean and 68\,\% confidence limits) on the parameters that we allow to vary in our analysis of the simulated data, along with their corresponding input values. }
\label{1dconstraints-sim}
\begin{tabular}{@{}lll@{}}
\toprule
Parameter                       & Constraints (68\,\% uncert.) & Input value \\ \midrule
$\Omega_{\mathrm{m}}$                      & $0.31 \pm 0.02$  & 0.315 \\
$\sigma_8$                      & $0.77 \pm 0.04$  & 0.811        \\
$h$                           & $0.67 \pm 0.02$ & 0.674   \\
$1-b_{\textnormal{SZ}}$         & $0.72 \pm 0.1$ & 0.62  \\
$1-b_{\textnormal{CMBlens}}$    & $0.91 \pm 0.05$  & 0.92 \\
$\sigma_{\textnormal{CMBlens}}$ & $0.23 \pm 0.05$ & 0.22  \\
$r_{\textnormal{CMBlens-SZ}}$   & $0.75 \pm 0.09$ & 0.77  \\ 
$\alpha$ & $1.84 \pm 0.06$ & 1.79  \\
$Y_{\star}$   & $0.65 \pm 0.03$ & 0.646  \\ \bottomrule

\end{tabular}
\end{table}

\section{Results and discussion}\label{results}

\begin{figure*}
\centering
\includegraphics[width=1.\textwidth]{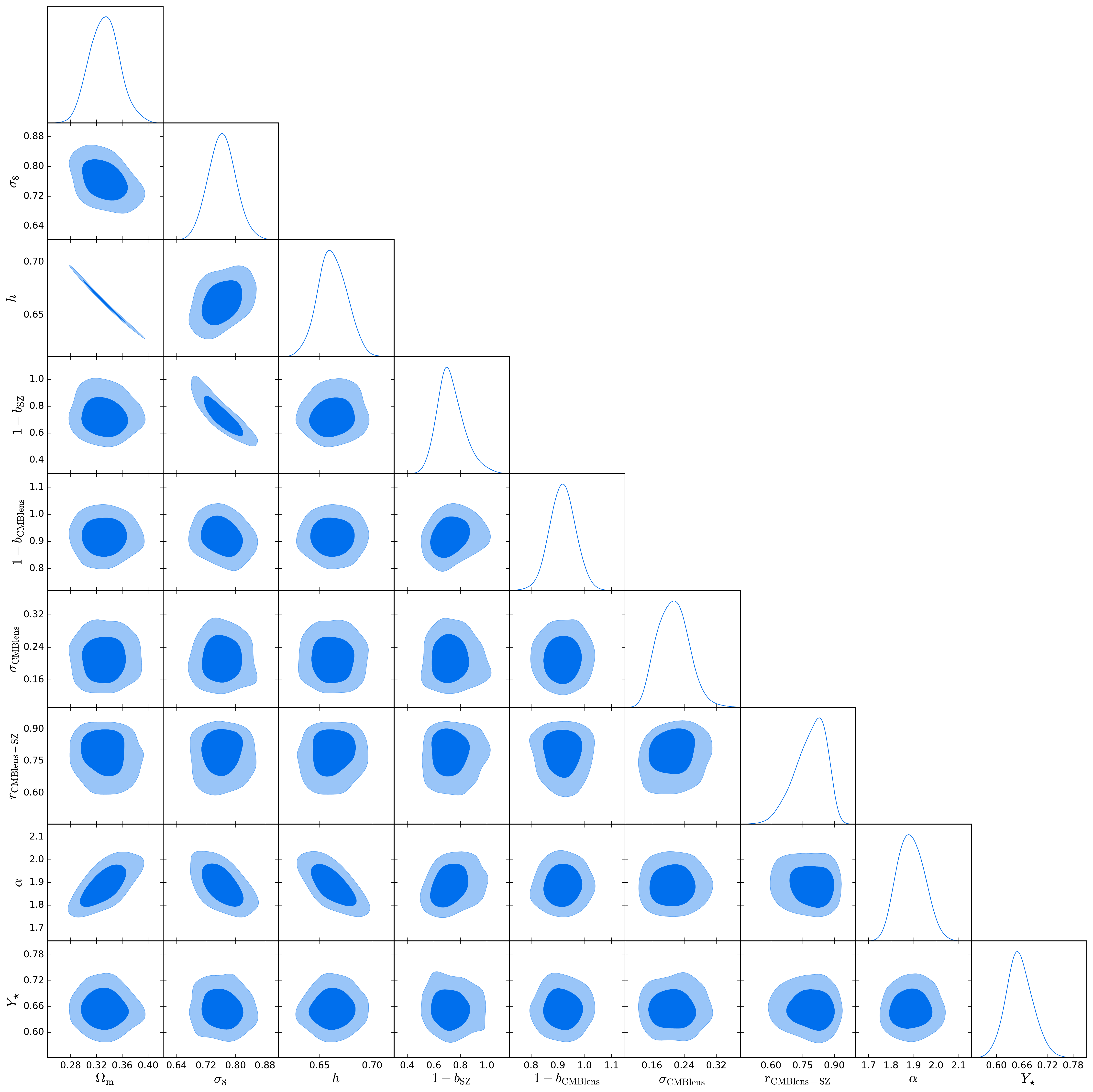}
\caption{Constraints (68\,\% and 95\,\% confidence regions) on the cosmological and cluster model parameters that we allow to vary in our analysis of the \textit{Planck} MMF3 cosmology sample of clusters with CMB lensing mass calibration and a prior on $\theta_{\mathrm{MC}}$. One-dimensional, marginalised posterior distributions are also shown.}
\label{corner_full}
\end{figure*}

\begin{figure}
\centering
\includegraphics[width=0.5\textwidth]{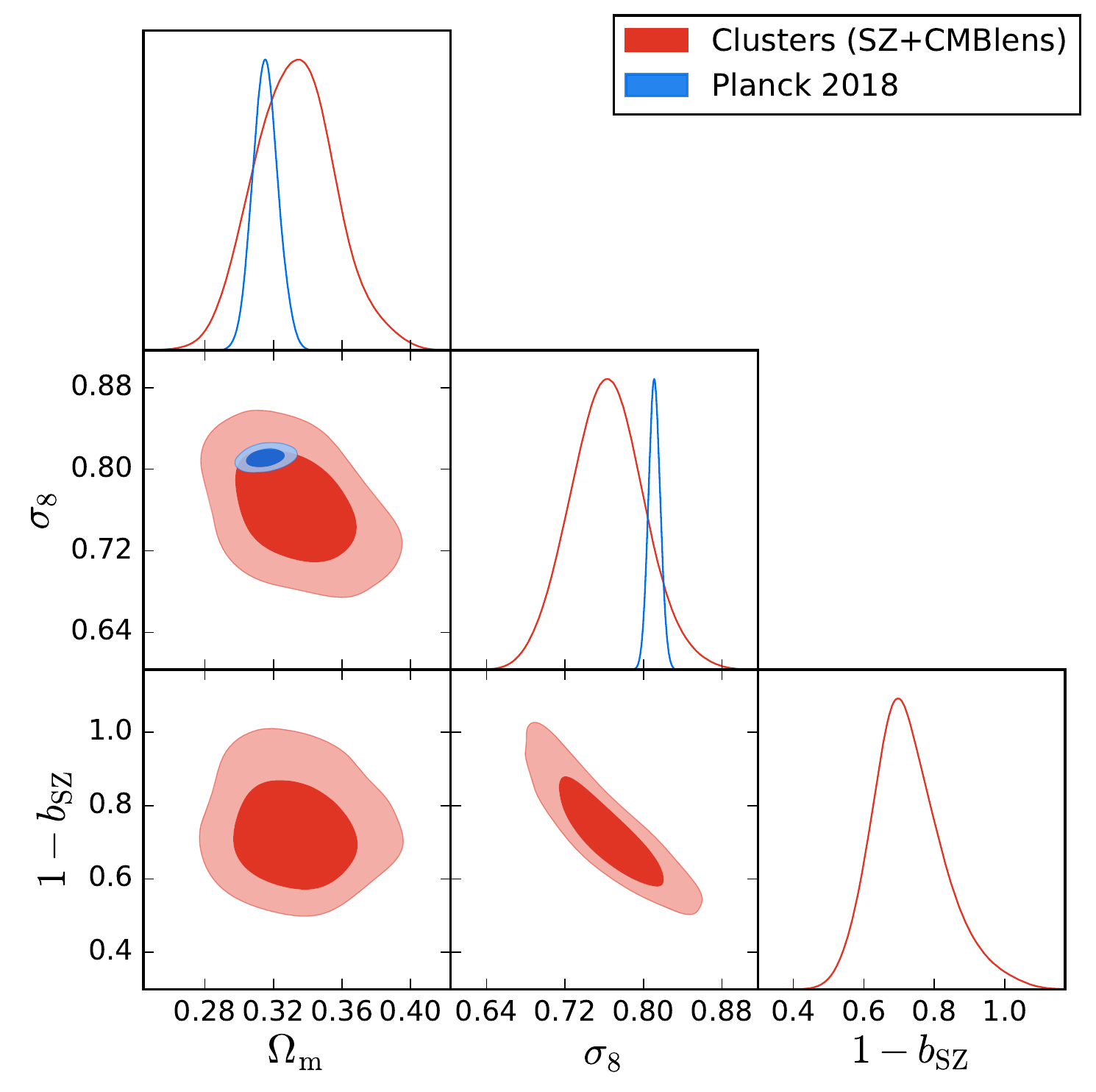}
\caption{Constraints on the cosmological parameters $\Omega_{\mathrm m}$ and $\sigma_8$, and on the SZ mass bias, $1-b_{\textnormal{SZ}}$, from our analysis of the \textit{Planck} MMF3 cosmology sample of clusters (red). We stress that the joint constraints on $\Omega_{\text{m}}$ and $\sigma_8$ along their long degeneracy axis are strongly dependent on the priors on $\alpha$ and $\beta$, whereas the perpendicular direction (the approximate combination $\sigma_8 \Omega_{\text{m}}^{0.25}$) is more immune to these priors. The one-dimensional constraints on $\Omega_{\text{m}}$ and $\sigma_8$ are similarly strongly affected. Also shown (in blue) are the constraints from the \textit{Planck} 2018 measurements of the CMB angular power spectra and CMB lensing, using the data combination TT,TE,EE+lowE+lensing~\citep{Planck2018}. Both constraints assume a flat $\Lambda$CDM cosmology.}
\label{corner_compare}
\end{figure}

We explore the posterior distribution for the real cluster data with an MCMC using the 
\texttt{emcee} package, as for our tests on simulated data. The two-dimensional marginalised constraints on all the parameters we allow to vary are shown in Fig.~\ref{corner_full}, and the two-dimensional marginalised constraints on $\Omega_{\mathrm{m}}$, $\sigma_8$, and $1-b_{\textnormal{SZ}}$ are shown in Fig.~\ref{corner_compare}, along with the corresponding constraints on the cosmological parameters from the \textit{Planck} 2018 measurements of the CMB angular power spectra and CMB lensing (the TT,TE,EE+lowE+lensing likelihood; \citealt{Planck2018}). In addition, the one-dimensional marginalised constraints on $\sigma_8 \left(\Omega_{\mathrm{m}}/0.33\right)^{0.25}$ and on our cluster parameters are given in Table~\ref{1dconstraints}.

We note that the constraints on $\Omega_{\mathrm{m}}$, $\sigma_8$, and $h$ are dependent on our choice of priors on $\alpha$ (Gaussian prior) and $\beta$ (delta function prior). As argued in Section \ref{subsec:paramsandpriors}, allowing $\beta$ to vary would widen the constraints in the $\Omega_{\mathrm{m}}$--$\sigma_8$ plane along the long degeneracy axis. Through the very strong degeneracy between $\Omega_{\mathrm{m}}$ and $h$ due to our prior on $\theta_{\mathrm{MC}}$, the constraints on $h$ would also widen accordingly. \citet{Ade2016} find that leaving $\alpha$ free also widens the constraints along the long degeneracy axis in the $\Omega_{\mathrm{m}}$--$\sigma_8$ plane; we expect a similar effect in our analysis, with constraints on $h$ also widening accordingly. We therefore decide to quote our cosmological constraints in terms of $\sigma_8 \left(\Omega_{\mathrm{m}}/0.33\right)^{0.25}$, a parameter that runs perpendicular to the long degeneracy axis in the $\Omega_{\mathrm{m}}$--$\sigma_8$ plane, and which therefore is more immune to the choice of priors on $\alpha$ and $\beta$. Purely for reference, we find $\Omega_{\mathrm{m}} = 0.33 \pm 0.02$, $\sigma_8 = 0.76 \pm 0.04$, and $h = 0.66 \pm 0.01$, stressing that they are dependent on the choice of priors on $\alpha$ and $\beta$.

\begin{table}
\centering
\caption{Marginalised constraints (mean and 68\,\% confidence limits) on $\sigma_8 \left(\Omega_{\mathrm{m}}/0.33\right)^{0.25}$ and on the cluster parameters that we allow to vary in our analysis of the \textit{Planck} MMF3 cosmology sample of clusters.}
\label{1dconstraints}
\begin{tabular}{@{}ll@{}}
\toprule
Parameter                       & Constraints (68\% uncert.) \\ \midrule
$\sigma_8 \left(\Omega_{\mathrm{m}}/0.33\right)^{0.25}$  & $0.765 \pm 0.035$ \\
$1-b_{\textnormal{SZ}}$         & $0.71 \pm 0.1$  \\
$1-b_{\textnormal{CMBlens}}$    & $0.92 \pm 0.05$  \\
$\sigma_{\textnormal{CMBlens}}$ & $0.21 \pm 0.04$  \\
$r_{\textnormal{CMBlens-SZ}}$   & $0.80^{+0.06}_{-0.09}$  \\
$\alpha$   & $1.89 \pm 0.07$  \\
$Y_{\star}$   & $0.65 \pm 0.03$   \\ \bottomrule
\end{tabular}
\end{table}
\begin{table}
\caption{Constraints (mean and 68\,\% confidence region) on $1-b_{\textnormal{SZ}}$ from the datasets indicated. These include weak lensing of background galaxies for subsets of \textit{Planck} clusters from Weighting the Giants (WtG;~\citealt{Linden2014}), the Canadian Cluster Comparison Project (CCCP;~\citealt{Hoekstra2015}), \citet[the ``cosmological subsample'']{Sereno2017}, \citet{Penna-Lima2017}, and \citet{Medezinski2018}. CMB-derived constraints are for
the CMB lensing calibration presented in \citet{Ade2016}, the \textit{Planck} 2015 SZ counts together with the \textit{Planck} 2018 TT,TE,EE+lowE likelihood \citep{Planck2018}, and this work. The constraint from the combined analysis of \textit{Planck} clusters, the tSZ power spectrum, and the tSZ bispectrum of \citet{Hurier2017} is also included.}
\label{bias_comparison}
\begin{threeparttable}
\centering
\begin{tabular}{ll}
\toprule
Dataset                                                      & $1-b_{\textnormal{SZ}}$ \\ \midrule
WtG                                                              & $0.688 \pm 0.072$      \\
CCCP                                                & $0.780 \pm 0.092$      \\
Sereno et al. (2017)\tnote{a} & $0.66\pm 0.10$     \\
Penna-Lima et al. (2017) & $0.73 \pm 0.10 $      \\
Medezinski et al. (2018) & $0.80 \pm 0.14 $ \\
Hurier et al. (2017) & $0.71 \pm 0.07$ \\
\textit{Planck} 2015 CMB lensing\tnote{b}      & $1.01^{+0.24}_{-0.16}$ \\
\textit{Planck} 2015 SZ + \textit{Planck} 2018 CMB & $0.62 \pm 0.04$        \\
This work                              & $0.71 \pm 0.10$ \\ \bottomrule
\end{tabular}
\begin{tablenotes}
\item[a] In our notation, \citet{Sereno2017} actually constrain $\ln(1-b_{\text{SZ}}) = -0.40\pm 0.14$.
\item[b] The CMB lensing measurement in \citet{Ade2016} is actually $1/(1-b_{\text{SZ}}) = 0.99 \pm 0.19$.
\end{tablenotes}
\end{threeparttable}
\end{table}

As it can be seen in Table \ref{1dconstraints}, our constraint on $\sigma_8 \left(\Omega_{\mathrm{m}}/0.33\right)^{0.25}$ from the \textit{Planck} MMF3 cosmology sample of clusters with CMB lensing calibration of the cluster masses
is consistent with that derived in the $\Lambda$CDM model from the \textit{Planck} 2018 CMB angular power spectra (and lensing), $\sigma_8 \left(\Omega_{\mathrm{m}}/0.33\right)^{0.25} = 0.8020 \pm 0.0085$. Consistency can also be seen in Fig.~\ref{corner_compare}. We therefore find no evidence for tension between the abundance of galaxy clusters measured at lower redshifts and the predictions for the $\Lambda$CDM model with parameters (mostly) calibrated at CMB decoupling. An alternative view of this consistency is to compare our constraint on the 
SZ mass bias parameter, $1-b_{\textnormal{SZ}} = 0.71 \pm 0.10$ (see Table~\ref{bias_comparison}), with that obtained in $\Lambda$CDM from joint analysis of the CMB angular power spectra and cluster counts without further mass calibration. The latter was reported as $1-b_{\textnormal{SZ}} = 0.62 \pm 0.04$ by~\citet{Planck2018}, showing good consistency.
 
Our CMB-lensing constraints on the SZ mass bias are also consistent with those from lensing of background galaxies by Weighing the Giants (WtG; \citealt{Linden2014}) and from the Canadian Cluster Comparison Project (CCCP; \citealt{Hoekstra2015}); see Table~\ref{bias_comparison}. These mass calibrations were used as priors in the analysis of \citet{Ade2016}. They are also consistent with the weak lensing constraints of subsamples of \textit{Planck} clusters by \citet{Penna-Lima2017} and \citet{Sereno2017}, with the constraint from the combined analysis of \textit{Planck} clusters and the tSZ power spectrum and bispectrum of \citet{Hurier2017}, and with the HSC galaxy weak lensing mass calibration of eight ACTPol clusters of \citet{Miyatake2018} and of five \textit{Planck} clusters of \citet{Medezinski2018} (see also Table~\ref{bias_comparison}). We note also that our constraint on $1-b_{\text{SZ}}$ is of comparable precision to current constraints from galaxy lensing. While the mass constraints on individual low-redshift clusters are weaker from current CMB lensing data than from galaxy lensing, the former approach can be applied to the full cluster sample.

Our constraint on the SZ mass bias, along with those from galaxy lensing, are in mild disagreement with that obtained from CMB lensing in \citet{Ade2016}. Reasons for this discrepancy were discussed in Section~\ref{naive}. We note that our constraint is, however, of similar precision: we find $1/(1-b_{\text{SZ}}) = 1.41 \pm 0.20$ from our samples of $b_{\text{SZ}}$.
% about a factor two more precise, despite using almost the same set of clusters (433 out of the 439 that we use in our analysis). \AC{Is this correct? With naive propagation of errors, I get $1/(1-b_{\text{SZ}}) = 1.41 \pm 0.20$ from the Bayesian constraint, which is very similar precision.} \IZL{Here I am comparing our calibration to the constraint from the \textit{Planck} CMB lensing calibration shown in Table 6 ($1-b_{\text{SZ}} = 1.01^{+0.24}_{-0.16}$). This value is reported in the \textit{Planck} 2015 SZ paper, in the `Notes' section of Table 2, as an approximate translation of their actual constraint on $1/(1-b_{\text{SZ}})$.}

%They remain, however, at a certain tension with the constraints from the CMB lensing calibration presented in \citet{Ade2016}. We also note that our constraints on $1-b_{\textnormal{SZ}}$ are competitive in precision with the constraints from galaxy lensing \citep{Linden2014,Hoekstra2015}, and about a factor of two more precise that the CMB lensing constraints presented in \citet{Ade2016}, in which almost the same set of clusters is used (433 out of the 439 that we use in our analysis).

There are some degeneracies between parameters that are worth commenting on.
The strongest degeneracy is that between $\Omega_{\mathrm{m}}$ and $h$ (see Figs.~\ref{corner_full} and~\ref{corner_compare}). This degeneracy is due to the prior we impose on $\theta_{\textnormal{MC}}$, which, at fixed baryon density, constrains $\Omega_{\text{m}}h^3$ to be approximately constant. More interesting, regarding the cluster counts, is the degeneracy between $1-b_{\textnormal{SZ}}$ and $\sigma_8$. As expected, there is a negative correlation between these parameters, which arises mostly through the dependence of our likelihood on the halo mass function. The physical interpretation of this degeneracy is straightforward: a given set of cluster SZ measurements can be explained with some given values of $1-b_{\textnormal{SZ}}$ and $\sigma_8$, or with a 
smaller value of $1-b_{\textnormal{SZ}}$ (making the mass larger at a given SZ cluster signal) and a larger value of $\sigma_8$ (to enhance the number of more massive clusters to match the observed number).
Finally, there is also an anti-correlation between $\Omega_{\mathrm{m}}$ and $\sigma_8$, arising primarily from the mass function: the increase in the number density of massive clusters with increasing $\sigma_8$ can be offset by the overall reduction in the number density of clusters with decreasing matter density.

Finally, we note that our CMB lensing signal-to-noise $\msn$ measurements are interesting per se in that they represent a significant detection of cluster CMB lensing with \textit{Planck} data. Since the observational scatter is much larger than the intrinsic scatter and than the scatter associated with the spread of $M_{500}$ and $z$ across the sample, our $\msn$ measurements roughly follow a Gaussian distribution with some mean and standard deviation of unity. If no cluster lensing signal were present, they would follow a Gaussian distribution with zero mean and unit standard deviation. We find $\left\langle \msn \right\rangle = 0.234 \pm 0.052$, where angular brackets denote averaging over the cluster sample, thus detecting the CMB cluster lensing signal at $4.5\,\sigma$ significance. For comparison, \citet{Baxter2015} detect the CMB lensing cluster signal of 513 SZ-selected SPT clusters with SPT data at $3.1\,\sigma$ significance, \citet{Baxter2017} detect that of 3697 high-redshift DES clusters with SPT data at $8.1\,\sigma$ significance, and \citet{Raghunathan2019} detect that of 4003 and 1741 DES clusters with SPTpol data at $8.7\,\sigma$ and $6.7\,\sigma$ significance, respectively. In addition, \citet{Madhavacheril2015} detect the CMB lensing signal of 12\,000 optically-selected CMASS galaxies using ACT data at $3.2\,\sigma$ significance.

\section{Conclusion}\label{conclusions}

In this paper we have presented constraints on the $\Lambda$CDM-model parameters $\Omega_{\mathrm{m}}$, $\sigma_8$, and $H_0$, and a CMB lensing calibration of the SZ mass bias, $1 - b_{\textnormal{SZ}}$, obtained from 439 SZ-selected galaxy clusters from the \textit{Planck} MMF3 cosmology sample. Our analysis revisits the \textit{Planck} SZ counts analysis with CMB lensing mass calibration presented in \citet{Ade2016}. The analysis there used cluster counts in the SZ signal-to-noise and redshift plane in order to constrain cosmological (and cluster model) parameters,
imposing a prior on $1 - b_{\textnormal{SZ}}$ derived from a CMB lensing mass calibration also presented in \citet{Ade2016}. Such calibration found no evidence for a SZ mass bias, in contrast to galaxy weak lensing calibrations on subsets of the cluster sample (e.g.,~\citealt{Linden2014,Hoekstra2015}) that favour $1-b_{\textnormal{SZ}}<1$ at more than $2\,\sigma$.  Although the statistical significance of the difference in mass bias measured from CMB and galaxy lensing is relatively weak, given the large measurement errors for CMB lensing, adopting the CMB lensing calibration of~\citet{Ade2016} exacerbates tension between cluster constraints in the $\sigma_8$--$\Omega_{\mathrm{m}}$ plane and those derived from the primary CMB anisotropies in $\Lambda$CDM.

We argued in Section~\ref{naive} that there are several effects that may have led to bias in this previous mass calibration from CMB lensing. We have remeasured the cluster masses via CMB lensing, and included the signal-to-noises of these and the SZ measurements, along with the cluster redshifts, in a Bayesian analysis that naturally takes account of all significant effects that likely biased the analysis in~\citet{Ade2016}. This approach allows us to constrain jointly the cosmological parameters and the SZ mass bias (and other cluster model parameters) in an unbiased way, as demonstrated through simulated data in Section~\ref{validation}.

%such a calibration is affected by several effects that may bias the estimate of the SZ mass bias (see Section \ref{naive}), and present an alternative approach in which the CMB lensing mass estimates of the clusters in our sample are incorporated into a likelihood that also takes into account the SZ signal-to-noise and the redshift measurements of each cluster. This likelihood naturally takes into account all the significant effects likely biasing the method followed in \citet{Ade2016}, and allows to constrain jointly $\Omega_{\mathrm{m}}$, $\sigma_8$, and $1 - b_{\textnormal{SZ}}$ in an unbiased way, as we verify in Section \ref{validation}.

With our likelihood, and including priors on some of the cluster model parameters informed by results from numerical simulations (but, notably, with no prior on the SZ mass bias $1-b_{\text{SZ}}$), we obtain constraints on $\Omega_{\mathrm{m}}$, $\sigma_8$, and $H_0$ that are consistent with the constraints in the $\Lambda$CDM model derived from the \textit{Planck} measurements of the CMB power spectra (see Figs.~\ref{corner_full} and \ref{corner_compare}, and Table \ref{1dconstraints}). We measure a significant SZ mass bias, $1-b_{\textnormal{SZ}} = 0.71 \pm 0.10$, consistent with measurements from galaxy weak lensing. We therefore find no evidence in our analysis of tensions between $\Lambda$CDM model parameters derived from the \textit{Planck} cluster sample and the primary CMB anisotropies.

Our work is further evidence of the growing power of CMB lensing to calibrate the overall cluster mass scale in SZ counts analyses. It will be a particularly powerful approach for future high-resolution CMB experiments, such as CMB-S4, which should detect around $10^5$ clusters through their SZ signatures \citep{Abazajian2016}. Indeed, CMB lensing allows one to estimate cluster masses from CMB data alone (plus, in principle, redshift measurements of each cluster), which implies that the whole cluster sample can be used in the calibration analysis, rather than just a small subsample, which is currently the case for galaxy lensing calibrations. It is also not affected by the uncertainties in the photometric redshifts of the background galaxies, which are a limiting factor in galaxy lensing analyses, and, furthermore, allows for the determination of masses of high-redshift clusters, where galaxy lensing mass reconstructions perform badly due to the dearth of background galaxies. For future CMB experiments such as CMB-S4, the CMB lensing signal will be such that, in SZ counts analyses, the SZ--mass scaling relations will be able to be calibrated completely with CMB lensing masses alone to sub-percent accuracy \citep{2017Louis}. In order to achieve this, however, work will be required in order to understand better the biases, intrinsic scatter, and correlations in the cluster observables, constraining them more accurately through numerical simulations and taking into account their dependence on mass and redshift.

\section*{Acknowledgements}

The authors would like to thank James G. Bartlett, Jean-Baptiste Melin, Sebastian Bocquet, and William Handley for useful discussions, and Anna Bonaldi and Richard Battye for providing us with the \textit{Planck} SZ filter noise estimates.

IZ is supported by the Isaac Newton Studentship from the University of Cambridge.

Our MCMC analyses were performed using the Cambridge Service for Data Driven Discovery (CSD3) operated by the University of Cambridge Research Computing Service (\url{http://www.csd3.cam.ac.uk/}), provided by Dell EMC and Intel using Tier-2 funding from the Engineering and Physical Sciences Research Council, and DiRAC funding from the Science and Technology Facilities Council (\url{www.dirac.ac.uk}).

%%%%%%%%%%%%%%%%%%%%%%%%%%%%%%%%%%%%%%%%%%%%%%%%%%


\bibliographystyle{mnras}
\bibliography{biblioteca} % if your bibtex file is called example.bib

%%%%%%%%%%%%%%%%%%%%%%%%%%%%%%%%%%%%%%%%%%%%%%%%%%

%%%%%%%%%%%%%%%%% APPENDICES %%%%%%%%%%%%%%%%%%%%%

\appendix

\section{Effect of miscentering on cluster mass measurements.}\label{appendix1}

In this appendix we provide further details of the bias caused by miscentering in our lensing cluster mass estimates. We also propose a method to account for miscentering bias more carefully in a Bayesian analysis of cluster data.

\subsection{Miscentering bias}

As discussed in Section~\ref{subsec:matchfilt}, we can build an estimator of a cluster's mass, $\hat{M}_{500}$, or, equivalently, of its signal-to-noise, $\msn$, by matched-filtering the cluster's reconstructed convergence profile with an appropriate template. This estimator is unbiased if the template matches exactly the true cluster's profile. This condition also requires that the template is placed exactly at the centre of the true profile. Here we consider the case in which the template is placed at a cluster's estimated centre, $\bmath{x}_{\textnormal{est}}$, which is offset from the true centre, $\bmath{x}_{\textnormal{true}}$, by a certain angle $\bmath{d}$, i.e., $\bmath{x}_{\textnormal{true}}  = \bmath{x}_{\textnormal{est}} + \bmath{d}$. Assuming that otherwise the template is exactly the same as the true cluster's profile, the expected value of $\msn$ can be written as
\begin{equation}
\left\langle \msn \right\rangle (\bmath{d}) = \mathcal{N}  \int \frac{d^2 \bmath{L}}{2 \pi} \frac{ | \kappa (\bmath{L})|^2  }{N_{\kappa}(\bmath{L})}e^{-i \bmath{L} \cdot \bmath{d}},
\end{equation}
where $\kappa$ is the true profile and $\mathcal{N}$ is an appropriate normalisation. Expanding the phase term in a power series, only the even powers of $\bmath{d}$ contribute since $\kappa(\bmath{L}) = \kappa^*(-\bmath{L})$ because the convergence is real. If we further assume circular symmetry for the convergence profile, so that $\kappa(\bmath{L})$ only depends on $L=|\bmath{L}|$, we can 
write the fractional bias due to miscentering by $\bmath{d}$ as
\begin{equation}\label{expansion}
   \frac{\langle \msn \rangle  (\bmath{d}) - \langle \msn \rangle  (\bmath{0})}{\left\langle \msn \right\rangle  (\bmath{0})} = \sum_{\text{$n$ even}} (-1)^{n/2}  f^{(n)} d^n .
\end{equation}
Here, the $f^{(n)}$ are coefficients quantifying the fractional bias at each order, which can be written as
\begin{equation}
   f^{(n)} = \frac{1}{2^n [(n/2)!]^2} \left(\int dL\, L \frac{ | \kappa (\bmath{L})|^2  }{N_{\kappa}(\bmath{L})}\right)^{-1}
\int dL\, L^{n+1} \frac{ | \kappa (\bmath{L})|^2  }{N_{\kappa}(\bmath{L})} .
\label{cluster}
\end{equation}
%
% %
% \begin{equation}
%    f^{(n)} = 
% \begin{cases}\label{cluster}
%     \frac{1}{n! (n+1)} \frac{\int dL \frac{ | \kappa (\bmath{L})|^2  }{N_{\kappa}(\bmath{L})} L^{n+2}}{\int dL \frac{ | \kappa (\bmath{L})|^2  }{N_{\kappa}(\bmath{L})} L^{2}} & \text{if } n \text{ even}, \\
%     0              & \text{if } n \text{ odd}. \\
% \end{cases}
% \end{equation}
% %
We see that the leading-order bias is second order in the miscentering angle and is negative (while the next-to-leading-order term is positive). 
% It can be seen that, due to the fact that the profile is circularly symmetric, there are only even contributions to the bias and they have an alternating sign. This is important, since it means that, for small miscentering angles, the bias is a second order effect.
We note that the fractional bias in $\msn$
is also the fractional bias in $\hat{M}_{500}$, since the latter is related to $\msn$ through a factor unaffected by miscentering.

To estimate the size of the effect of miscentering, we can write the $n$th order contribution to the bias as
\begin{equation}
(-1)^{n/2} f^{(n)} d^n = (-1)^{n/2} \left( \frac{d}{\lambda^{(n)}} \right)^n,
\end{equation}
where $\lambda^{(n)}$ is a typical angle given by $\lambda^{(n)} = ( f^{(n)})^{-1/n}$. For a \textit{TT} quadratic estimator reconstruction in an idealised \textit{Planck}-like experiment with a Gaussian beam of $\text{FWHM} = 5\,\text{arcmin}$ and temperature noise levels of $45\,\text{$\mu$K-arcmin}$, we find $\lambda^{(2)} \approx 6$\,arcmin and $\lambda^{(4)} \approx 7$\,arcmin. For a cluster with $M_{500}=0.5 \times 10^{15} M_{\odot}$ at $z=0.3$ and a typical miscentering error of around $1$\,arcmin, the second-order fractional bias is around $10^{-2}$ (i.e., at the few percent level), while the fourth-order bias is around $10^{-4}$.

\subsection{Accounting for miscentering in a likelihood analysis}

As explained in Section~\ref{bayesian}, in our likelihood analysis we model the miscentering bias with an effective lensing mass bias parameter, which we can marginalise over. Here we describe a more rigorous way to deal with miscentering bias in a likelihood analysis, although we leave implementation of this approach to future work.

If we knew the centering offset $\bmath{d}$, for a cluster with true lensing signal-to-noise $p_{\text{t}}$ the average of the observed signal-to-noise over other sources of observational scatter is
\begin{equation}
\langle \msn \rangle(\bmath{d}) \approx p_{\text{t}} \left(1+\sum_{\text{$n$ even}} (-1)^{n/2}  f^{(n)} d^n\right).
\label{debiasmiscent}
\end{equation}
Here, the bias coefficients $f^{(n)}$ should be evaluated at the appropriate cluster parameters ($M_{500}$, $z$, $\hat{\bmath{n}}$, $1-b_{\text{CMBlens}}$) in the hierarchical model. However, they are independent of $\bmath{d}$, which proves useful for efficient marginalisation over $\bmath{d}$.

Obviously, $\bmath{d}$ is not known precisely, but experiments such as \textit{Planck} do provide an estimate of the uncertainty in each reported cluster position. For a given cluster, let us assume that its position on the sky is described by a certain two-dimensional probability distribution, $P(\bmath{d})$. This could be a two-dimensional Gaussian, for example.  What we are interested in is the probability distribution followed by $\msn$, $P(\msn|p_{\text{t}},M_{500},z,\hat{\bmath{n}})$, after the uncertainty due to miscentering has been taken into account. This is given by
\begin{equation}
P(\msn|p_{\text{t}},M_{500},z,\hat{\bmath{n}}) = \int d^2 \bmath{d} \, P(\msn|p_{\text{t}},M_{500},z,\hat{\bmath{n}},\bmath{d})
P(\bmath{d}) ,
\label{eq:Pmsnint}
\end{equation}
where $P(\msn|p_{\text{t}},M_{500},z,\hat{\bmath{n}},\bmath{d})$ is the conditional distribution for $\msn$ given the centering error $\bmath{d}$. Retaining the Gaussian approximation for the other observational sources of scatter, this conditional distribution is simply a Gaussian with mean given by Eq.~\eqref{debiasmiscent} and unit variance.

The series expansion in Eq.~\eqref{debiasmiscent} allows $P(\msn|p_{\text{t}},M_{500},z,\hat{\bmath{n}})$ to be evaluated efficiently.
We note that if indeed we take $P(\msn|p_{\text{t}},M_{500},z,\hat{\bmath{n}},\bmath{d})$ and $P(\bmath{d})$ both to be Gaussians, and if we correct for miscentering at least to the lowest non-vanishing order (quadratic), then $P(\msn|p_{\text{t}},M_{500},z,\hat{\bmath{n}})$ will no longer be Gaussian and will depend (weakly) on $M_{500}$, $z$, and $\hat{\bmath{n}}$ through the $f^{(n)}$.
Thus, the method outlined here allows to account for miscentering uncertainty in a rigorous way at the expense of complicating the statistics of our mass observable. We note that a similar miscentering argument should apply to the SZ signal-to-noise measurements, which should have common miscentering errors with the CMB lensing signal-to-noise measurements. Accounting properly for miscentering would therefore imply marginalising over $\bmath{d}$ simultaneously in both observables.

\section{Relating our likelihood to a Poisson counts likelihood}\label{appendix3}

In this appendix we demonstrate how the likelihood used in this paper, which is described in detail in Section~\ref{bayesian}, is mathematically equivalent to a Poisson counts likelihood in the $q_{\mathrm{obs}}$--$p_{\mathrm{obs}}$--$z$--$\hat{\bmath{n}}$ space for cells of sufficiently small size. We also show how it can be reduced to a likelihood similar to the one used in the \textit{Planck} SZ counts analysis \citep{Ade2016} by suitable marginalisation.

Let us consider an $M$-dimensional data space for each galaxy cluster; in our case, $M=5$. We continue to assume that clusters can be treated independently of each other (i.e., effects of spatial clustering are negligible), so that our cluster sample can be considered a realisation of a Poisson process in this data space. One way to describe the sample statistically is via a Poisson counts approach. 

In this approach, we first divide the $M$-dimensional (continuous) data space into a set of $N_{\textnormal{bin}}$ cells, covering the whole region of data space where there is a non-zero probability for a cluster to be located. We assume that each cell has equal volume $\Omega$.
Since the clusters are assumed to be statistically independent from each other, the number of clusters with data falling within cell $I$, $n_I$, follows a Poisson distribution, and the joint probability distribution for the sample, now described by the cell occupation numbers $\bmath{n}=(n_1,\ldots,n_{N_{\text{bin}}})$ is simply given by the product of the Poisson distributions of each cell,
\begin{equation}
P(\bmath{n}) = \prod_I \frac{\bar{n}_I^{n_I} e^{-\bar{n}_I}}{n_I!}.
\end{equation}
Here, $\bar{n}_I$ is the expected number of clusters in cell $I$. We can write $\bar{n}_I = P_I \bar{N}$, where $P_I$ is the probability for a single cluster to lie in cell $I$ given that it is passes the selection criterion to be in the sample, and $\bar{N}=\sum_I \bar{n}_I$ is the expected total number of clusters in the sample.
We can then rewrite the likelihood as
\begin{equation}
P(\bmath{n}) = \frac{\bar{N}^N e^{-\bar{N}}}{N!} N! \prod_I \frac{P_I^{n_I}}{n_I!},
\label{eq:Poissonfac}
\end{equation}
where $N=\sum_I n_I$ is the total number of clusters in the sample.
The first term in Eq.~(\ref{eq:Poissonfac}) is a Poisson distribution for $N$, which we shall refer to as $P(N)$. The second term is a multinomial distribution for $\bmath{n}$ with $N$ trials, that is, the probability distribution followed by $\bmath{n}$ given a fixed $N$. We shall refer to it as $P(\bmath{n} | N)$. We have therefore factored a product of Poisson distributions across the cells into a Poisson distribution for the total number of clusters and a distribution for the particular configuration of cluster data points at a fixed total number of clusters,
\begin{equation}
P(\bmath{n}) = P(\bmath{n} | N) P(N).
\end{equation}

Let us now consider how this Poisson counts likelihood relates to the unbinned likelihood used in the rest of this paper. Let us denote points in the $M$-dimensional data space by $\bmath{x}$ and the data corresponding to the $i$th cluster as $\bmath{x}_i$. Also, let $P(\bmath{x}_i|\inc)$ be the probability density for a cluster to have data $\bmath{x}_i$ given that it passes the sample-selection criterion. A given realisation of the cluster sample consists of specifying $N$ and the set $\underline{D}'=(\bmath{x}_1,\ldots,\bmath{x}_N)$. As the cluster labels are arbitrary, there are $N!$ permutations that give equivalent samples of clusters. Each permutation gives the same cell occupation numbers $\bmath{n}$, which are got from counting the number of clusters $n_{I}$ that have their data $\bmath{x}_i$ within cell $I$. Now consider the limit in which the cells that are small enough that the variation of $P(\bmath{x}|\inc)$ across the cell can be ignored. If some central value of $\bmath{x}$ in the $I$th cell is denoted by $\bmath{x}_I$, for a given sample of $N$ clusters we have
\begin{align}
\prod_{I=1}^{N_{\text{bin}}}  P_I^{n_I} &\approx \prod_{I=1}^{N_{\text{bin}}}  \left[P(\bmath{x}_I|\inc) \Omega \right]^{n_I} \nonumber \\
&\approx \prod_{i=1}^N \left[P(\bmath{x}_i|\inc) \Omega \right] \nonumber \\
&= P(\underline{D}'|N) \Omega^N .
\end{align}
Thus, in this limit, we can rewrite the cells likelihood, Eq.~(\ref{eq:Poissonfac}), as
\begin{equation}\label{decom}
P(\bmath{n}) \propto P(\underline{D}'|N) P(N),
\end{equation}
where the proportionality factor, which accounts for the change in measure between the two data descriptions, is cosmology independent.

Equation~(\ref{decom}) has the same form as our likelihood, presented in Section \ref{bayesian}. There, we separated the angular positions of the $N$ clusters, $\underline{\hat{\bmath{n}}}$, from the other cluster observables, $\underline{D}$, so that $\underline{D}' = (\underline{D},\underline{\hat{\bmath{n}}})$. Indeed, Eq.~(\ref{likelihoodfull}) can be written as
\begin{equation}
P\left( N, \underline{\hat{\bmath{n}}}, \underline{D} \right) = P(\underline{D}, \underline{\hat{\bmath{n}}} | N ) P(N) = P(\underline{D}'|N)P(N) ,
\end{equation}
where $P(\underline{D}, \underline{\hat{\bmath{n}}} | N )$ is just the product of the mass data likelihood, $\mathcal{L}_1$ (see Section~\ref{clusterdata}), and the sky location likelihood, $\mathcal{L}_2$ (see Section~\ref{subsec:skylocL}).
Our likelihood is therefore equivalent to a Poisson counts likelihood in the $D$--$\hat{\bmath{n}}$ space (or, more explicitly,
the $q_{\mathrm{obs}}$--$p_{\mathrm{obs}}$--$z$--$\hat{\bmath{n}}$ space), with cells of sufficiently small size. This also implies that if we were to marginalise our likelihood over all the $p_{\mathrm{obs}}$ and $\hat{\bmath{n}}$ variables, we would obtain a likelihood equivalent to a Poisson counts likelihood in the $q_{\mathrm{obs}}$--$z$ plane (with cells of sufficiently small size). This is similar to the \textit{Planck} SZ counts likelihood used in \citet{Ade2016}, except there cells with a fixed finite size are used.

%%%%%%%%%%%%%%%%%%%%%%%%%%%%%%%%%%%%%%%%%%%%%%%%%%

% Don't change these lines
\bsp	% typesetting comment
\label{lastpage}
\end{document}